\documentclass[twocol]{ametsoc}

\journal{jpo}

\bibpunct{(}{)}{;}{a}{}{,}
\def\citeapos#1{\citeauthor{#1}'s [\citeyear{#1}]}

\newcommand{\figref}[1]{Fig.\ \ref{#1}}

\def\pct{\%{ }}

\def\const{\mathop{\rm const}}

\title{Statistics of simulated and observed pair separations in
the Gulf of Mexico}

\authors{Francisco J.\ Beron-Vera\correspondingauthor{Francisco J.\
Beron-Vera, RSMAS/ATM, University of Miami, 4600 Rickenbacker Cswy.,
Miami, FL 33149.}}

\affiliation{Department of Atmospheric Sciences, Rosenstiel School
of Marine and Atmospheric Science, University of Miami, Miami,
Florida, USA.}

\email{fberon@rsmas.miami.edu}

\extraauthor{J. H.\ LaCasce} 

\extraaffil{Department of Geosciences, University of Oslo, Oslo, Norway.}

\abstract{Pair-separation statistics of in-situ and synthetic surface
drifters deployed near the \emph{Deepwater Horizon} site in the
Gulf of Mexico are investigated.  The synthetic trajectories derive
from a 1-km-resolution data-assimilative Navy Coastal Ocean Model
(NCOM) simulation.  The in-situ drifters were launched in the Grand
LAgrangian Deployment (GLAD).  Diverse measures of the dispersion
are calculated and compared to theoretical predictions. For the
NCOM pairs, the measures indicate nonlocal pair dispersion (in which
pair separations grow exponentially in time) at the smallest sampled
scales.  At separations exceeding 100 km, pair motion is uncorrelated,
indicating absolute rather than relative dispersion. With the GLAD
drifters however the statistics are ambiguous, with some indicating
local dispersion (in which pair separations exhibit power law growth)
and others suggesting nonlocal dispersion.  The difference between
the two data sets stems in part from inertial oscillations, which
affect the energy levels at small scales without greatly altering
pair dispersion.  These were significant in GLAD but much weaker
in the NCOM simulation.  In addition the GLAD drifters were launched
over a limited geographical area, producing few independent
realizations and hence lower statistical significance.  Restricting
the NCOM set to pairs launched at the same locations yields very
similar results, suggesting the model is for the most part capturing
the observed dispersion.}

\begin{document}

\maketitle

\section{Introduction}

Submesoscale processes, i.e., with length scales of 0.1--10 km
\citep{Thomas-etal-08}, are believed to be important in the upper
ocean \citep[][]{McWilliams-08a, Klein-09}.  These are the transition
scales between the largely balanced quasi-2D flows at the mesoscales
and 3D (unbalanced) flows at smaller scales.  While observational
evidence of submesoscale activity in the ocean is accumulating,
important questions about their dynamics and the consequences for
transport remain.

In the atmosphere, the balanced scales are characterized by a kinetic
energy spectrum proportional to $k^{-3}$, where $k$ is the horizontal
wavenumber \citep{Nastrom-Gage-85}.  The consensus is that this
reflects a quasi-2D enstrophy cascade toward smaller scales
\citep{Kraichnan-67, Charney-71}. At scales of several hundred
kilometers, the spectrum transitions to a $k^{-5/3}$ dependence.
This primarily reflects divergent motions (inertia--gravity waves),
at scales where the Rossby number exceeds one \citep{Callies-etal-14}.
\citet{Callies-Ferrari-13} suggest a similar situation exists in
the ocean.

The slope of the energy spectrum is important for Lagrangian transport
\citep[e.g.,][]{Bennett-06, LaCasce-08}. With a $-5/3$ slope the
dispersion of pairs of particles (or ``relative dispersion'') is
\emph{local}, meaning separations between pairs of particles are
dominated by eddies of comparable scales. With a $-3$ or steeper
slope the dispersion is \emph{nonlocal} and governed by the largest
eddies in the $k^{-3}$ range.  Local dispersion results in small
scale ``billowing,'' as with smoke from a stack, while nonlocal
dispersion produces filaments.  Particle dispersion can thus be
used to infer aspects of the energy spectrum, which can be useful
in the ocean at scales below those resolved by satellite altimetry.

Hereafter we examine relative dispersion at the surface of the Gulf
of Mexico (GoM).  The study was motivated by the Grand LAgrangian
Deployment (GLAD), which was conducted in the vicinity of the
\emph{Deepwater Horizon} (DwH) site in July 2012 and in which a
large number of surface drifters were deployed \citep{Olascoaga-etal-13,
Poje-etal-14, Jacobs-etal-14, Coelho-etal-15}.  A primary goal of
GLAD was to study dispersion at the submesoscales in the GoM.

Relative dispersion in the GoM has been studied previously.
\citet{LaCasce-Ohlmann-03} examined ``chance pairs'' of drifters
(i.e., drifters not deployed together) from the Surface-CUrrent and
Lagrangian drifter Program (SCULP) \citep{Ohlmann-Niiler-05} and
found nonlocal dispersion below the deformation radius, $L_\mathrm{D}$,
which is approximately 45 km in the GoM \citep{Chelton-etal-98}.
Supporting evidence, using pair separation probability distribution
functions (PDFs), was obtained by \citet{LaCasce-10}.  However,
using different measures (the second order longitudinal velocity
structure function and the separation-averaged relative diffusivity)
with the GLAD drifters, \citet{Poje-etal-14} concluded the dispersion
was local, from few hundred meters to several hundred kilometers,
implying a shallower kinetic energy spectrum.

Relative dispersion is often studied using two types of measures
\citep{LaCasce-08}. The first treats \emph{time} as the independent
variable. This includes the relative dispersion (the mean square
pair separation), the kurtosis (the normalized fourth moment) and
the separation PDF. The second type uses \emph{distance} as the
independent variable. This includes the structure functions, the
separation-averaged relative diffusivity and the finite-scale
Lyapunov exponent (FSLE) \citep{Artale-etal-97, Aurell-etal-97}.
The two types often produce different results, but the reasons for
this are rarely examined.

In the present paper, we examine both types of statistics, using
synthetic particles and drifters in the GoM.  The former were
obtained by integrating surface velocities produced by a data-assimilative
simulation with the 1-km-resolution Navy Coastal Ocean Model (NCOM)
\citep{Jacobs-etal-14}, and the latter are the drifters in the GLAD
experiment. The model allows for large numbers of particles,
increasing statistical reliability, whereas the drifters more
accurately reflect the actual situation in the GoM.

The paper is organized as follows.  In Section 2 and Appendix A we
present relevant theory for pair-separation statistics.  In Section
3 we examine the NCOM pair separations, and the GLAD pairs in Section
4.  A summary and concluding remarks are offered in Section 5.
Details of the numerical simulation and the GLAD experiment are
given in Appendices B and C, respectively.

\section{Theory}\label{sec:theory}

Let $r_0$ be the distance between two fluid parcels at time $t =
0$ and $r$ be the separation at time $t$.  In homogeneous, stationary,
and isotropic 2D turbulence, the PDF of pair separations, $p(r,t)$,
obeys a Fokker--Planck equation:
\begin{equation}
  \partial_t p = r^{-1} \partial_r(r\kappa_2 \partial_r p),
  \label{eq:fp}
\end{equation}
where $\kappa_2(r)$ is the scale ($r$) dependent relative diffusivity.
The 3D version of \eqref{eq:fp} was proposed by \citet{Richardson-26}
to describe smoke dispersion in the atmospheric boundary layer.
Richardson obtained a self-similar solution, based on an empirical
diffusivity derived from observations.  The equation was later
derived by \citet{Kraichnan-66} using his ``abridged Lagrangian
history direct interaction approximation'' and by \citet{Lundgren-81},
assuming an advecting velocity with a short correlation time. For
an overview, see \citet[Chapter 11]{Bennett-06}.

Pair dispersion depends on whether the pair velocities are correlated
or not, specifically whether the normalized Lagrangian velocity
correlation
\begin{equation}
  \frac{2\langle v_i \cdot v_j \rangle}{\langle v_i^2 \rangle +
  \langle v_j^2 \rangle} = 1 - \frac{\langle (v_i - v_j)^2
  \rangle}{\langle v_i^2 \rangle + \langle v_j^2 \rangle} 
  \label{eq:corr}
\end{equation}
(where the angle bracket indicates statistical average) equals 1
or 0, respectively \cite[e.g.,][]{Koszalka-etal-09}. 
The second term on 
the right side is proportional to the 
second-order velocity structure function,
\begin{equation}
  S_2(r) := \langle v^2 \rangle \equiv \langle (v_i - v_j)^2
  \rangle, 
  \label{eq:s2}
\end{equation}
where $v$ is the difference in the Eulerian velocity between points
separated by a distance $r$.  The Eulerian--Lagrangian equivalence
in \eqref{eq:s2} is a distinguishing aspect of homogeneous, isotropic
turbulence \citep{Bennett-84}.

At large separations, when the pair velocities are uncorrelated,
the relative diffusivity $\kappa_2$ is constant and equal to twice
the single particle diffusivity. At smaller scales, $\kappa_2$ can
be inferred if the energy spectrum has a power law dependence, i.e.,
$E(k) \propto k^{-\alpha}$ \citep{Bennett-84}.  In either case
equation (\ref{eq:fp}) can then be solved \citep{Bennett-06}.
Relevant 2D solutions are given in \citet{LaCasce-10} and
\citet{Graff-etal-15}, and are reproduced in Appendix A. From these,
the (raw) statistical moments, given by
\begin{equation}
  \langle r^n \rangle := 2\pi \int_0^\infty
  r^{n+1}p(r,t)\,\mathrm{d}r,
  \label{eq:mom}
\end{equation}
can be calculated.  Table \ref{tab:sum} shows the time dependences
for $\langle r^2 \rangle$ (the relative dispersion) and the $\langle
r^4 \rangle/\langle r^2 \rangle^2$ (kurtosis) in the three specific
2D dispersion regimes considered here.

\begin{table}
  \caption{The pair-separation measures in the three regimes
  considered here.  The measures are the kinetic energy spectrum,
  $E(k)$, where $k$ is wavenumber; the two particle diffusivity,
  $\kappa_2(r)$, where $r$ denotes separation; the second-order
  longitudinal structure function, $S_2(r)$; the relative dispersion,
  $\langle r^2 \rangle$; and the separation kurtosis, $\langle r^4
  \rangle/\langle r^2 \rangle^2$. The latter two are based on the
  (raw) statistical moments of the time-dependent probability
  distribution function (PDF) of pair separations, obeying \eqref{eq:fp}.
  The results shown for the Richardson and Rayleigh cases are the
  asymptotic (long time) limits.} \label{tab:sum}
  \begin{center}
  \begin{tabular}{lcccc}
  \topline & Lundgren & Richardson & Rayleigh \\ \midline $E(k)$ &
  $\propto k^{-3}$ & $\propto k^{-5/3}$  & --   \\ $\kappa_2(r)$ &
  $=r^2/T$ & $=\beta r^{4/3}$ & $= \const$ \\ $S_2(r)$      &
  $\propto r^2$ & $\propto r^{2/3}$  & $= \const$ \\ $\langle r^2
  \rangle$ & $= r_0^2 \mathrm{e}^{8t/T}$ & $\sim 5.2675 \beta^3
  t^3$  & $\sim 4 \kappa_2 t$ \\ $\langle r^4 \rangle/\langle r^2
  \rangle^2$ & $= \mathrm{e}^{8t/T}$ & $\sim 5.6$ & $\sim 2$ \\
  \botline
  \end{tabular}
  \end{center}
\end{table}

The nonlocal regime [which we refer to as the ``Lundgren regime,''
after \citet{Lundgren-81}] corresponds to an energy spectrum at
least as steep as $k^{-3}$, with a structure function $S_2 \propto
r^2$.  The PDF is not self-similar, but rather becomes more and
more peaked. Both the dispersion and kurtosis increase exponentially
in time.  The local regime we will consider has a Kolmorogorov
energy spectrum, $E \propto k^{-5/3}$, or equivalently $S_2 \propto
r^{2/3}$, and referred as the ``Richardson regime'' after
\citet{Richardson-26}.  In this case the separation PDF asymptotes
to a self-similar form, with a kurtosis of 5.6, and the dispersion
increases as time cubed.  With uncorrelated pair velocities, the
second-order structure function is constant with separation (and
equal to twice the mean square single particle velocity). The PDF
also asymptotes to a self-similar form, with kurtosis of 2 and the
dispersion increasing linearly in time. This self-similar PDF is a
Rayleigh distribution, so we refer to this as the ``Rayleigh regime.''

\section{Simulated pair-separation statistics}

\begin{figure}
  \centerline{%
  \includegraphics[width=.475\textwidth]{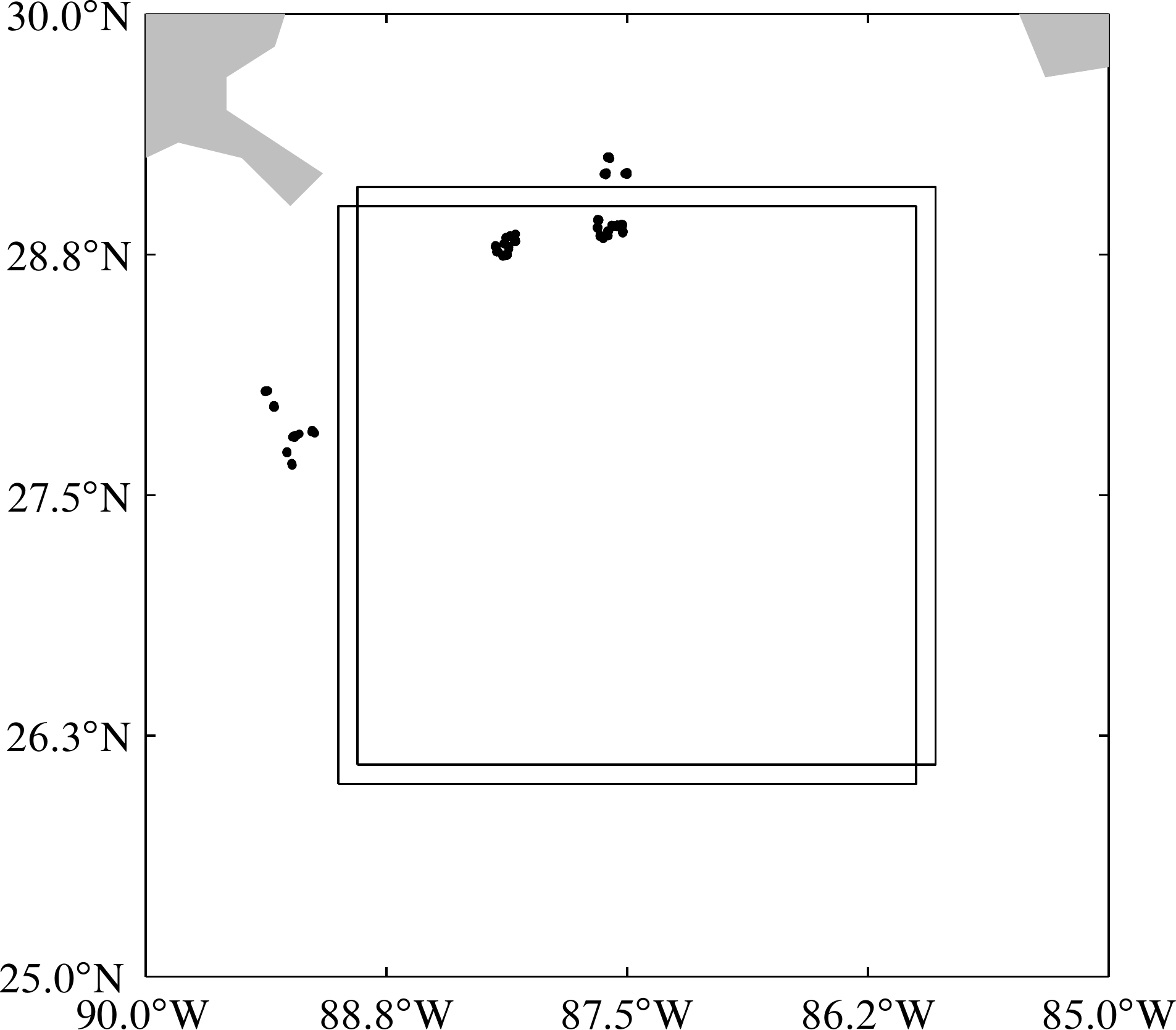}}% 
  \caption{The rectangles indicate the boundaries of the grids of
  initial positions for the integration of trajectory pairs (with
  $r_0 = 1$ km initial separation) using surface velocities produced
  by the 1-km-resolution Navy Coastal Ocean Model (NCOM) simulation
  of the Gulf of Mexico (GoM).  Dots indicate deployment locations
  of drifters from the Grand LAgrangian Deployment (GLAD).}
  \label{fig:ics}%
\end{figure}

The simulated trajectories were constructed by integrating surface
velocities produced by an NCOM simulation (cf.\ Appendix B).  The
integrations were carried out using a stepsize-adapting fourth/fifth-order
Runge--Kutta method with interpolations obtained using a cubic
scheme.  One-month-long records, with 10 positions per day, were
produced with a range of initial separations, from the smallest
scale resolved by the model simulation up to 30 km ($r_0 =$ 1, 5,
10 and 30 km).  The trajectories were initiated every other day in
the northern GoM near the DwH site, in two $5 \times 5$ 100-km-width
grids displaced by the chosen separation.  The reference and auxiliary
grids with $r_0 = 1$ km are shown in \figref{fig:ics}.  The
trajectories were started 1 July 2013 and 1 February 2014 to survey
summer and winter conditions.  The two were expected to exhibit
different pair-separation statistics, as the mixed layer is deeper
in winter.

\begin{figure}
  \centerline{%  
  \includegraphics[width=.475\textwidth]{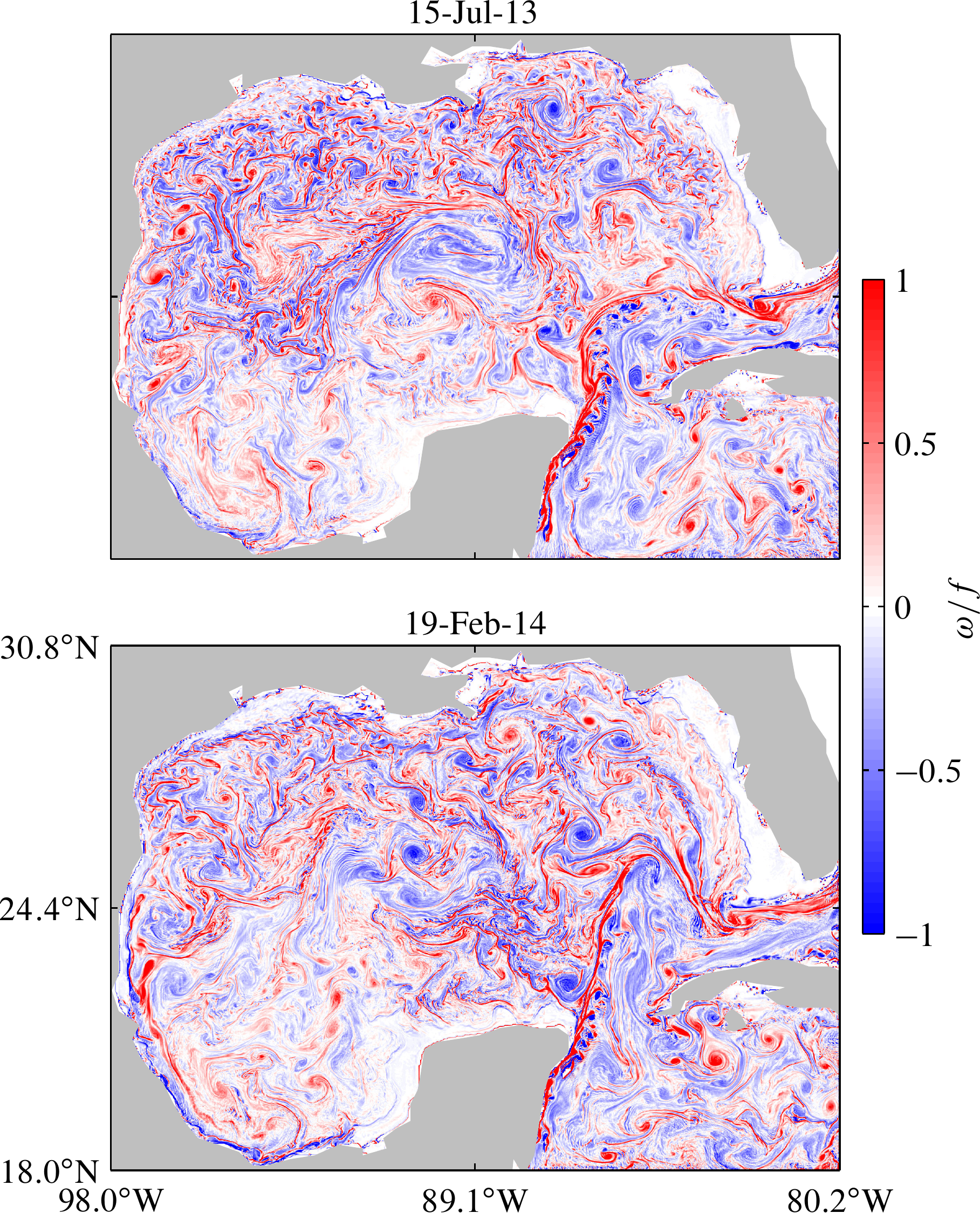}}%
  \caption{Snapshots of surface vorticity (normalized by
  the mean Coriolis parameter in the GoM) from the NCOM simulation
  in summer (top) and winter (bottom).}
  \label{fig:vor}%
\end{figure}

However, snapshots of the instantaneous surface vorticity
(\figref{fig:vor}) reveal roughly the same range of eddy scales in
the two seasons. The results seen hereafter similarly show only
small changes with season.  These figures also reveal that the West
Florida Shelf and the Bay of Campeche are relatively eddy inactive
regions, and that an anticyclonic ring has pinched off from the
Loop Current. Apart from these regions however the eddy field does
not exhibit significant spatial variability.  Thus we assume
homogeneity holds fairly well, as assumed in Section
\ref{sec:theory}.

An additional theoretical prerequisite is stationarity.  While
temporal variability of the simulated background eddy field is
evident, this mainly manifests on seasonal timescales.  Therefore,
considering motion over a period of 1 month, as we do here, is not
restrictive but rather necessary for stationarity to be fairly well
guaranteed.

\begin{figure}
  \centerline{%
  \includegraphics[width=.475\textwidth]{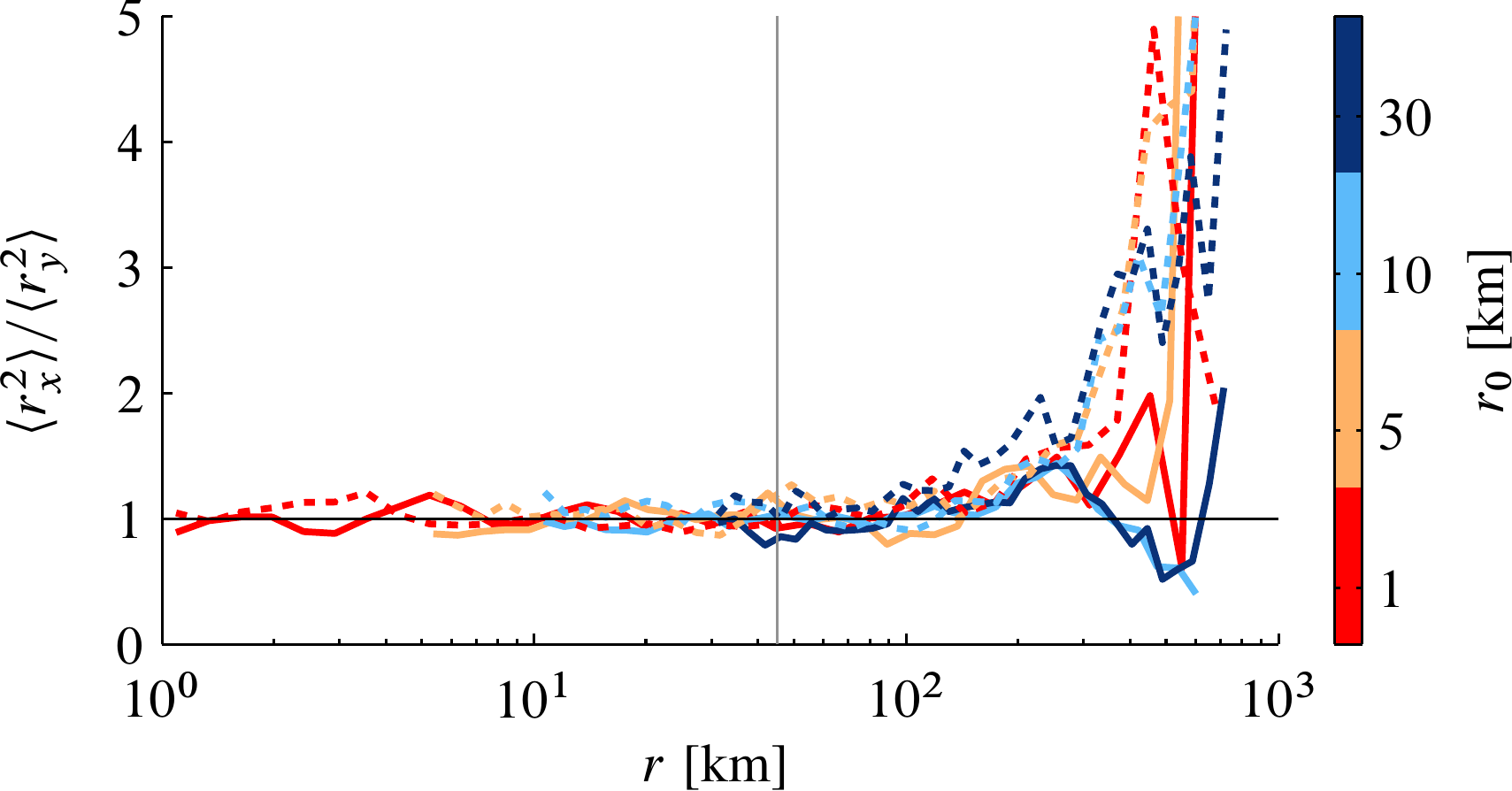}}%
  \caption{Zonal-to-meridional relative dispersion ratio as a
  function of pair separation in summer (solid) and winter (dashed)
  based on simulated trajectory pairs with various initial separations.
  The vertical line indicates the gravest baroclinic Rossby deformation
  radius.}
  \label{fig:iso}%
\end{figure}

Isotropy, the remaining prerequisite, is also realized. This can
be seen by plotting the ratio of the zonal to the meridional relative
dispersion ratio as a function of scale \citep{Morel-Larcheveque-74},
as shown in \figref{fig:iso}.  Isotropy holds in both summer (solid)
and winter (dashed), irrespective of the initial pair separation.
At separations exceeding 100 km, the dispersion becomes zonally
anisotropic, as it does in the atmosphere \citep{Graff-etal-15}.
But below that, the dispersion is isotropic.

\begin{figure}
  \centerline{%
  \includegraphics[width=.475\textwidth]{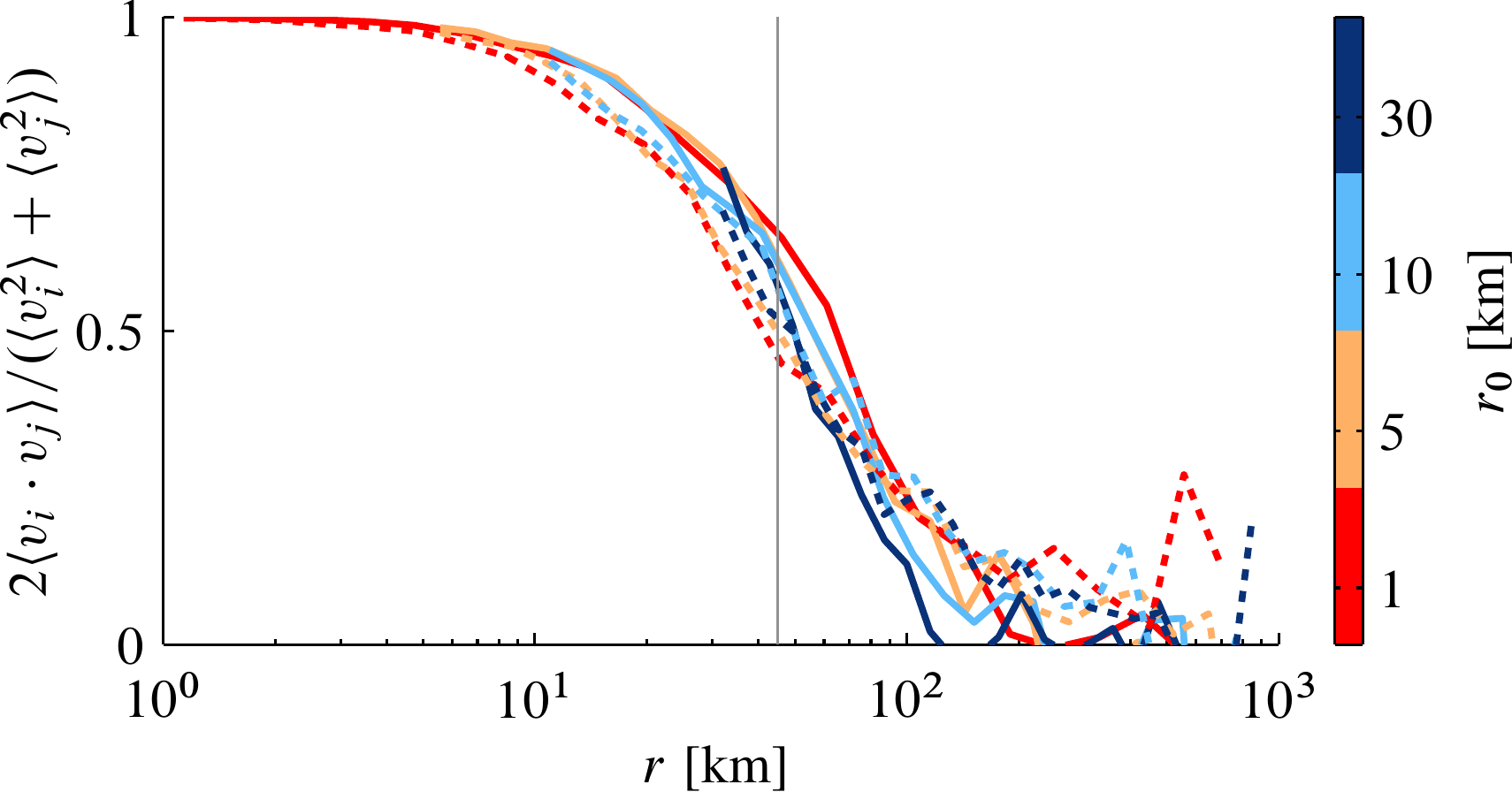}}%
  \caption{The normalized Lagrangian velocity correlation as function
  of scale in summer (solid) and winter (dashed) for various initial
  separations in the NCOM simulation.  The vertical line indicates
  the Rossby deformation radius.}
  \label{fig:cor}%
\end{figure}

We then determine over which scales the pair motion is correlated.
Inspection of \figref{fig:cor} reveals that, fairly insensitive to
the initial pair separation, the motion is correlated below the
deformation radius, in both seasons. At $L_\mathrm{D}$, the correlation
is roughly 0.5. The motion is strongly correlated below 20 km and
decorrelated above 100 km. Thus the proper framework for interpretation
above 100 km is absolute dispersion, possibly with the inclusion
of a mean zonal shear.

\begin{figure}
  \centerline{%
  \includegraphics[width=.475\textwidth]{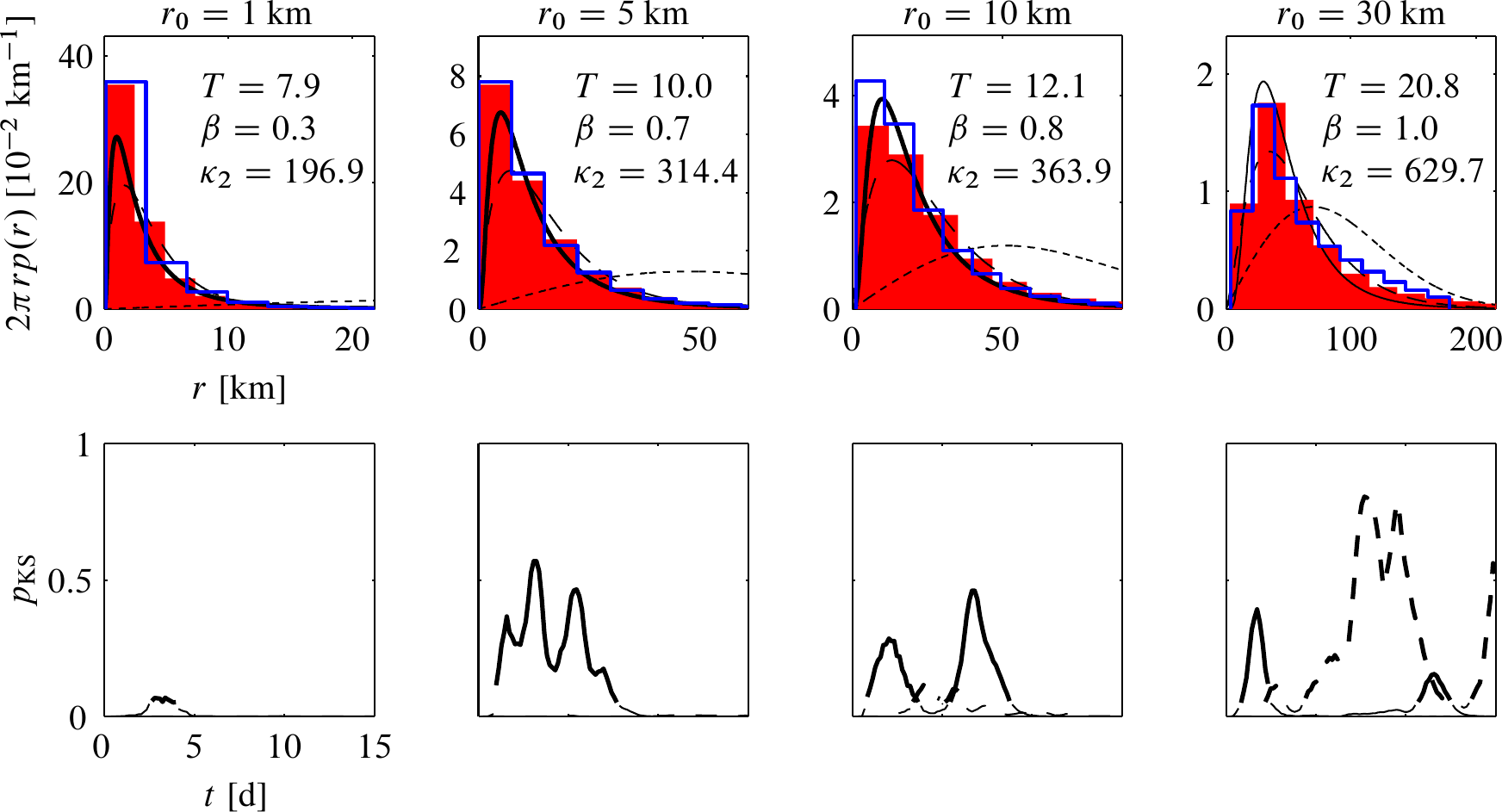}}%
  \caption{(top) Pair separation probability distribution
  functions (PDFs) in summer (red) and winter (blue) for various
  initial separations.  The PDFs are shown at time $t = 3.5$ d.
  Solid and dashed, and dot-dashed curves are theoretical PDFs
  in the Lundgren, Richardson, and Rayleigh regimes,
  respectively. Parameters $T$ (with units d$^{-1}$) and $\beta$
  (with units km$^{2/3}$d$^{-1}$) for the Lundgren and
  Richardson PDFs, respectively, were obtained by comparing the
  theoretical and observed dispersion curves via least squares
  out to the time by which $\sqrt{\langle r^2\rangle} = 5r_0$.
  The relative diffusivity $\kappa_2$ (with units
  km$^2$d$^{-1}$) for the Rayleigh distribution was calculated
  as the mean of the relative dispersion divided by $4t$ after
  $t = 10$ d (Appendix A).  When a theoretical curve is
  statistically similar to the observed PDF in summer at the
  95\pct confidence level, it is drawn in bold.  (bottom) The
  Kolmogorov-Smirnov probability for the theoretical curves vs
  the summer PDF as a function of time.}
  \label{fig:pdf}%
\end{figure}

We now inspect the pair separation PDFs and compare them to the
theoretical predictions (\figref{fig:pdf}). We calculated the PDFs
using kernel density estimation as implemented in Matlab's function
\texttt{ksdensity} with positive support.  Each theoretical PDF has
two parameters: the initial separation, $r_0$, and a growth parameter.
Since all pairs have the same initial separation, we assume $r_0$
is the same as in the simulation.  The growth parameters for the
Lundgren and Richardson distributions ($T$ and $\beta$, respectively)
were determined by fitting the dispersion using least squares during
the initial period, up to the point when the root-mean-squared (rms)
separation was a factor $a$ greater than the initial value.  We
chose $a = 5$, but the results are fairly insensitive to the choice.
The parameter for the Rayleigh PDF ($\kappa_2$) was calculated from
the relative dispersion at late times, after the pair motion was
decorrelated, as described in Appendix A.

The PDFs are plotted with the theoretical curves at $t = 3.5$ d.
Several initial separations are used, for particles deployed in summer
(in red) and winter (in blue), and the parameters obtained for the
theoretical curves are indicated in the inserts.  The observed and
theoretical PDFs were compared using the Kolmogorov--Smirnov (KS)
statistic \cite[cf., e.g.,][]{Press-etal-07}; since the winter and
summer distributions are so similar, we used the summer PDF for
this. The degrees of freedom for the KS test are determined by the
number of independent pairs. The pairs were deployed every 2 d and
100-km apart in the present simulation, so we can safely treat all
pairs as independent.  In the figure, the bold curve is statistically
similar to the observed summer PDF, with 95\pct confidence. The lower
panels show how the KS probability evolves with time, over the first
15 days. When a curve exceeds 0.05, it is plotted in bold.

Consider the case with $r_0 = 1$ km initial separation. The summer
and winter PDFs are very similar and are highly kurtosed, with most
pairs having small separations but some having much larger ones.
The Lundgren distribution (solid) has a similar shape, and indeed
is statistically the same at the 95\pct confidence level.  In
contrast, the Richardson (long-dashed) and Rayleigh (short-dashed)
distributions have a shorter tail and much larger mode, respectively.
The KS probability (lower panel) though suggests the Lundgren PDF
is significantly similar only during a brief period, near $t = 3$
d.

The results with $r_0 = 5$ and $10$ km are similar, in that observed
PDFs resemble the Lundgren distribution. The similarity moreover is
significant at the 95 \pct level for much of the first 10 d. The
Richardson PDF on the other hand is not similar over the same period.
With $r_0 = 30$ km, the PDF is statistically similar to the Lundgren
only briefly, but then resembles the Richardson. However, recall
that 30 km is only slightly less than the deformation radius, when
the pair velocities are significantly decorrelated. We do not find
similarity with the Rayleigh distribution in any case, during these
initial 15 d. 

\begin{figure}
  \centerline{%
  \includegraphics[width=.475\textwidth]{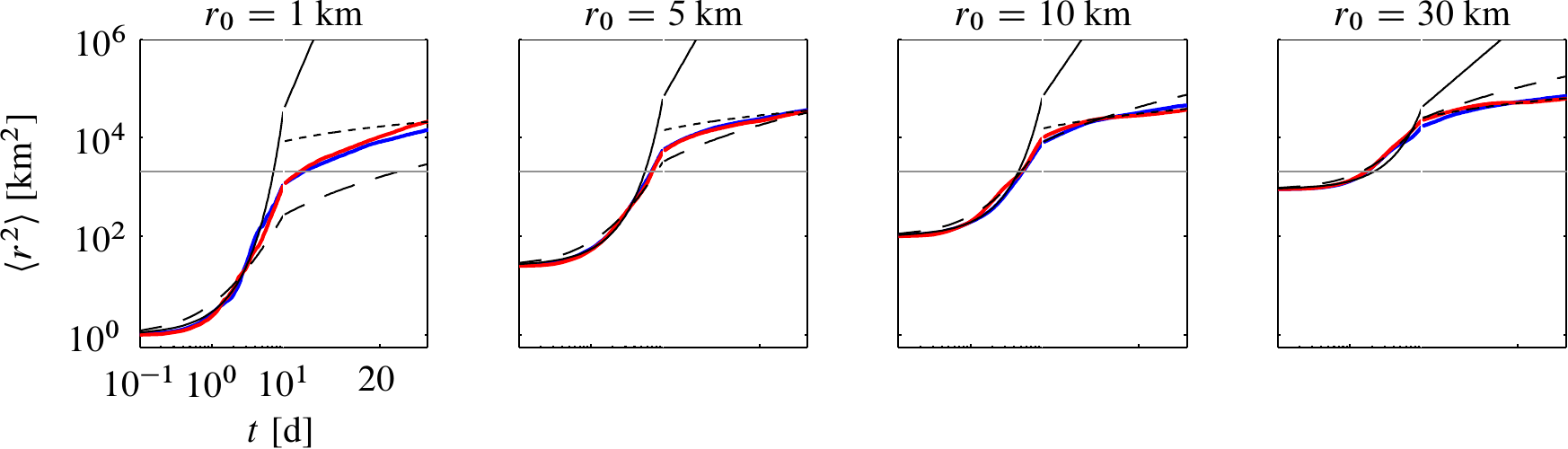}}%
  \caption{Relative dispersion (second moment of the pair-separation
  PDF) in summer (red) and winter (blue) for the NCOM pairs with
  various initial separations.  The Lundgren and Richardson predictions
  are indicated in solid and long-dashed curves, respectively.  The
  short-dashed line is the long-time asymptotic Rayleigh relative
  dispersion.  The horizontal dashed line indicates the deformation
  radius. Note that the $t$-axis is logarithmic until $t = 10$ d
  and linear thereafter.}
  \label{fig:r2}%
\end{figure}

The relative dispersion curves are shown in \figref{fig:r2}. The
winter (red) and summer (blue) curves are nearly identical, again
supporting similar behavior in the two seasons. The initial growth is
close to exponential initially, with an e-folding time on the order of
1 d. The exception is the $r_0 = 30$ km case, where the separations
are only briefly below the deformation radius (indicated by the
horizontal line). With the two larger initial separations, the late
dispersion increases linearly, in line with diffusive growth, but this
is not apparent in the 1 and 5 km cases. There is a clear suggestion
of Richardson-like growth with $r_0 = 10$ km, but in the other cases
the dispersion is either greater or less than this.

\begin{figure}
  \centerline{%
  \includegraphics[width=.475\textwidth]{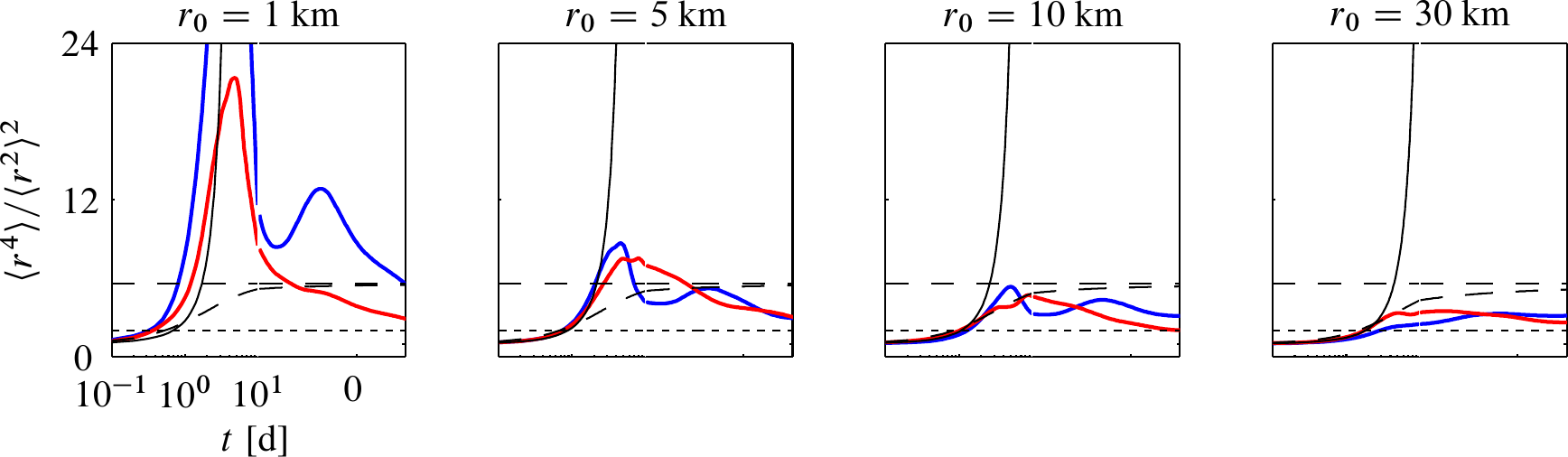}}%
  \caption{Kurtosis (normalized fourth moment of the pair-separation
  PDF) in summer (red) and winter (blue) for the NCOM pairs with
  various initial separations.  The Lundgren and Richardson curves
  are indicated by the solid and long dashed curves, respectively.
  Parameter-independent long-time asymptotic limits in the Rayleigh
  and Richardson regimes are indicated by the short- and long-dashed
  horizontal lines. As in the previous figure, the $t$-axis is
  logarithmic until $t = 10$ d and linear thereafter.}
  \label{fig:r4}%
\end{figure}

Being the fourth moment of the PDF, the kurtosis (\figref{fig:r4}) is
more sensitive to the tails of the distribution. With the smallest
initial separation ($r_0 = 1$ km), the kurtosis grows rapidly,
reaching values greater than 20, in both seasons.  Under nonlocal
dispersion the kurtosis grows exponentially and at the same rate as
the dispersion (Table \ref{tab:sum}). The observed growth is roughly
consistent with this, with an e-folding time of roughly 1 d.  With
$r_0 = 5$ km, the initial growth is curtailed and the maximum values
obtained are less. With $r_0 = 10$ and 30 km, the kurtosis quickly
relaxes toward 2, the asymptotic limit for the Rayleigh
distribution. There is little support for a Richardson regime here;
the kurtosis exceeds the asymptotic limit of 5.6 at the smallest
separations and falls below that at the larger separations.

\begin{figure}
  \centerline{%
  \includegraphics[width=.475\textwidth]{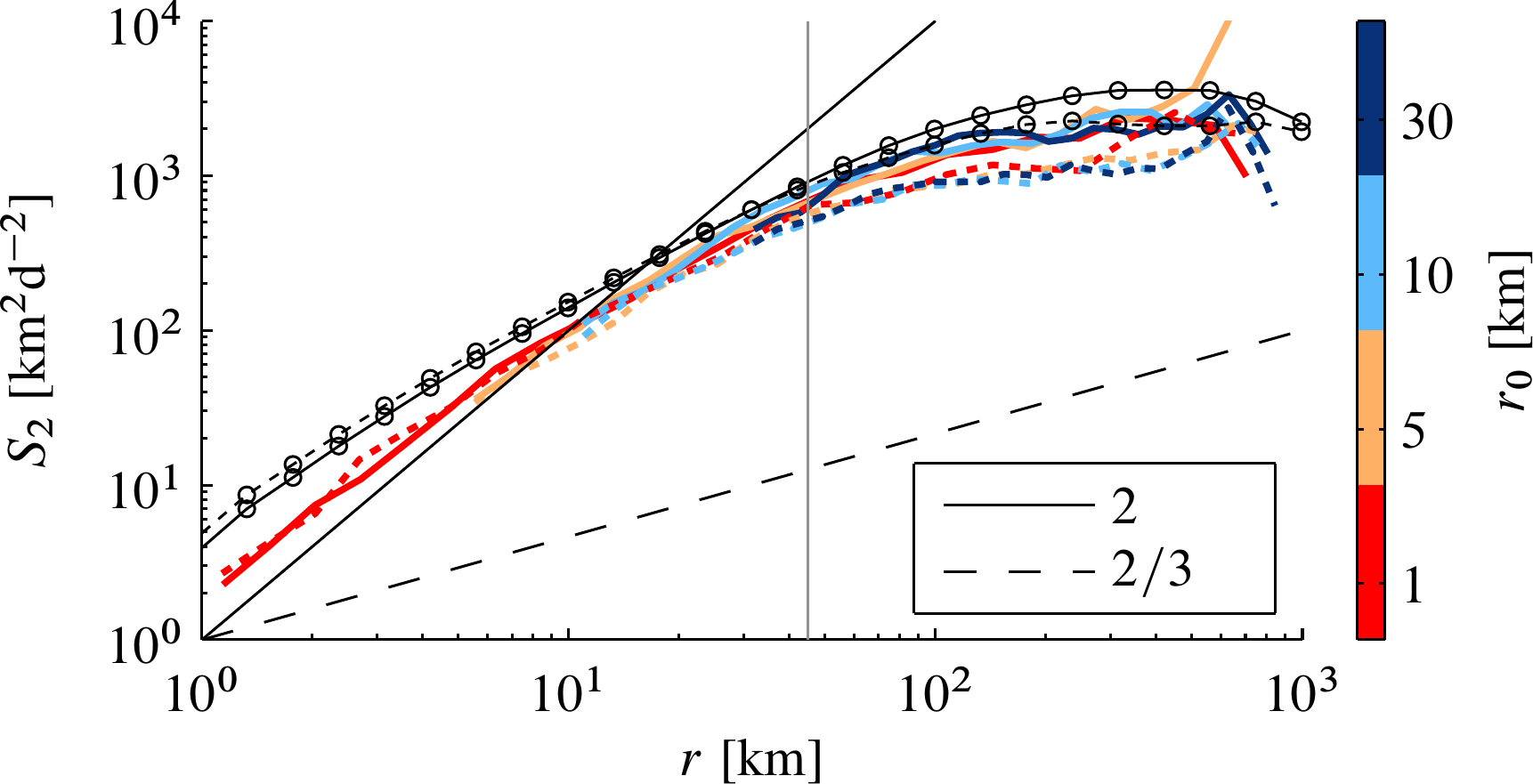}}%
  \caption{Second order structure functions in summer (solid) and
  winter (dashed). Colored curves are Lagrangian estimates for
  different initial separations.  The curves with circles are
  Eulerian estimates.  Solid and dashed straight lines have structure
  function slopes (indicated) in the theoretical Lundgren and
  Richardson regimes, respectively. The vertical line indicates the
  Rossby deformation radius.}
  \label{fig:s2}%
\end{figure}

The velocity structure functions ($S_2$) are shown in \figref{fig:s2}
for summer (solid) and winter (dashed) pairs.  These are very
similar, flattening out at scales exceeding 100 km, as expected for
uncorrelated motion.  With $r_0 = 1$ km, $S_2$ exhibits the $r^2$
dependence expected in a nonlocal regime at the smallest separations.
With larger $r_0$ the curves do not have a unique power-law-dependence
but suggest instead a transition between the $r^2$ and $r^0$
asymptotic limits. We also calculated the Eulerian $S_2$ using
velocity time series at fixed positions (curves with circles), and
this mirrors the Lagrangian $S_2$ well, supporting the assumptions
of homogeneity and isotropy.

Thus both the time- and distance-averaged measures indicate nonlocal
dispersion at the smallest separations and uncorrelated motion above
100 km. There was some suggestion of an intermediate Richardson regime
(primarily in the dispersion), but the kurtoses and structure
functions imply these scales are rather a transition between the small
and large scale limits.

One wonders of course to what extent numerical dissipation in the
model is responsible for the nonlocal dispersion at the smallest
scales.  So we turn to the GLAD drifters, which are not so affected.

\section{Observed pair-separation statistics}

The GLAD pairs were obtained from quarter-hourly drifter positions
from the GLAD experiment (cf.\ Appendix C).  The drifters were
deployed near the DwH site, as indicated by the black dots in
\figref{fig:ics}.  Various initial separation classes were identified:
$r_0 \approx$ 0.15, 1, and 10 km. A total of 132, 127, and 276
original pairs were obtained in each class. \footnote{The specific
initial separation ranges are: $0.12\text{ km} < r_0 < 0.19\text{
km}$, $0.84\text{ km} < r_0 < 1.19\text{ km}$, and $9.85\text{ km}
< r_0 < 10.19\text{ km}$.}  As in \citet{Poje-etal-14}, we consider
trajectory records spanning the initial 25 d after deployment, to
avoid enhanced windage effects on the drifters during the passage
of hurricane \emph{Issac}.

\begin{figure}
  \centerline{%  
  \includegraphics[width=.475\textwidth]{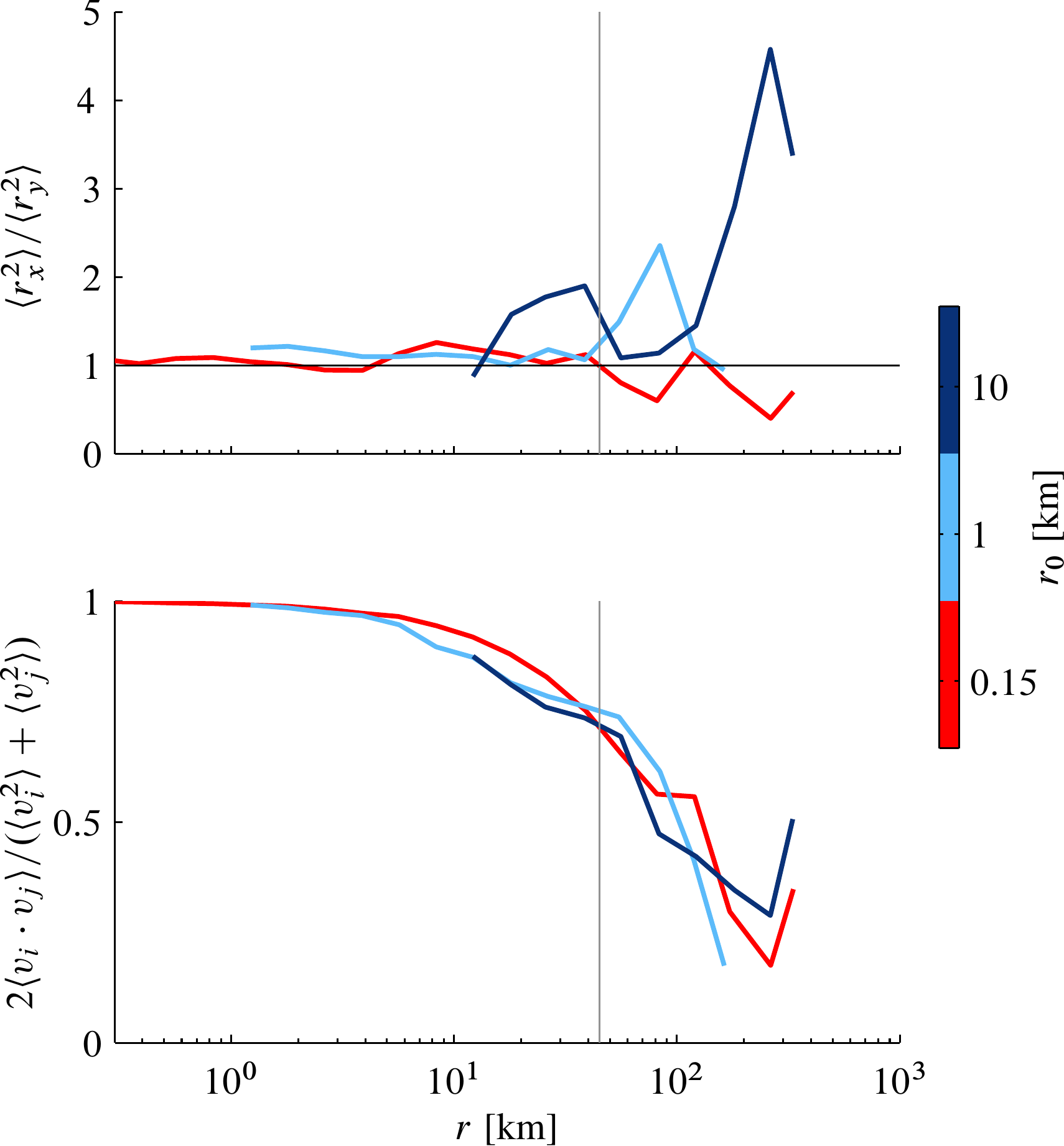}}%
  \caption{As in Figs.\ \ref{fig:iso} (top) and \ref{fig:cor}
  (bottom) but based on GLAD trajectory pairs in three 
  initial separation classes.}
  \label{fig:isocor-glad}%
\end{figure}

As the trajectories span no more than one month, we assume stationarity
holds. Homogeneity cannot be determined with the available data,
but the NCOM results suggest this is not an unreasonable assumption.
Isotropy is found for separations less than about 100 km or so
(\figref{fig:isocor-glad}, top panel). Furthermore, the pair
velocities are correlated over the isotropic scales, with correlations
falling below 0.5 above 100 km (\figref{fig:isocor-glad}, bottom
panel).

\begin{figure}
  \centerline{%
  \includegraphics[width=.475\textwidth]{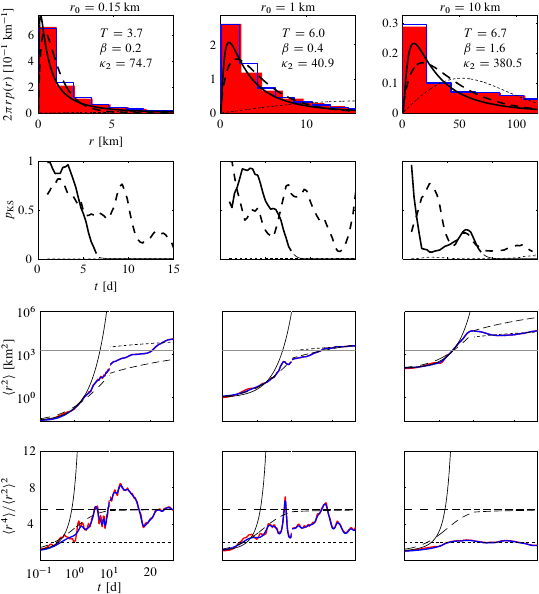}}%
  \caption{As in Figs.\ \ref{fig:pdf} (first and second rows),
  \ref{fig:r4} (third row), and \ref{fig:r2} (fourth row) but for
  separations experienced by the raw (red) and lowpass-filtered (blue)
  GLAD trajectory pairs.  The cut-off period for the filter is 2.5 d.}
  \label{fig:pdfr2r4-glad}%
\end{figure}

The time-based statistics are shown in \figref{fig:pdfr2r4-glad}.
The parameters for the theoretical curves were obtained as before,
by matching the observed dispersion from the initial separation up
to the scale at which the rms separation $\sqrt{\langle r^2 \rangle}
= 5 r_0$. In principle one could also treat the initial separation
as a free parameter, since a range of initial values is present,
but we chose to set $r_0$ equal to the mean value for the drifters
in each chosen range.

The GLAD PDFs 3-d after deployment are shown in red in the top panels
of \figref{fig:pdfr2r4-glad} for the 3 initial separations above.  The
PDFs are somewhat coarser than for the synthetic particles as there
are fewer pairs, but they too are peaked at the smallest separations
and exhibit extended wings.

Again, assessing the differences with the theoretical PDFs requires
knowing the number of degrees of freedom, i.e., the number of
independent realizations.  However this number was fairly small for
the GLAD experiment, as the deployment was made over a limited
geographical area.  Drifter clusters were deployed very near one
another, so that the distance between different clusters was often
much less than the putative energy-containing eddy scale of 100 km.
As such, many of the pair trajectories are similar. 

\begin{figure}
  \centerline{%
  \includegraphics[width=.475\textwidth]{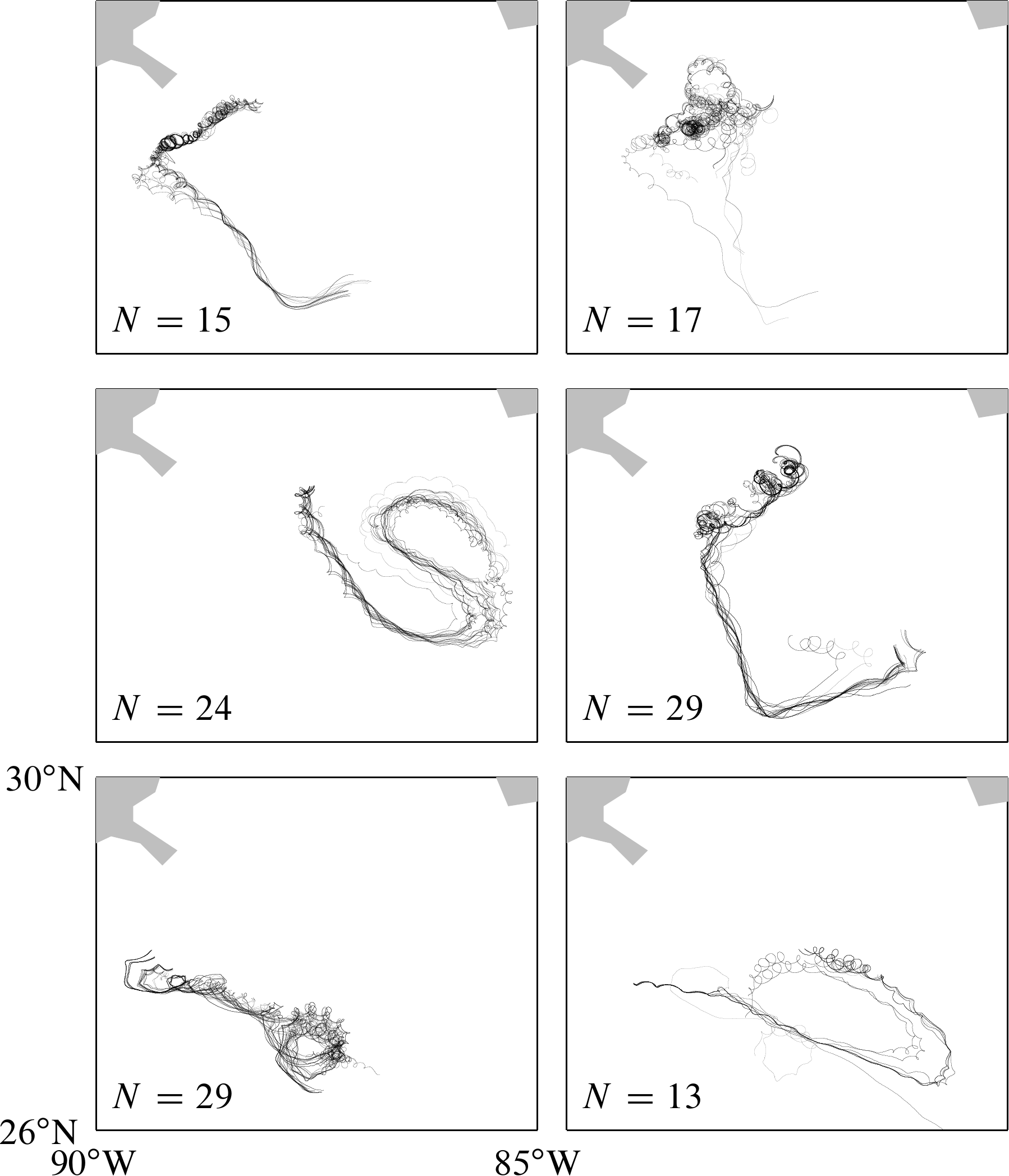}}%
  \caption{GLAD trajectory pairs with initial separation $r_0 \approx
  1$ km arranged into groups exhibiting similar behavior.  The
  number ($N$) of pairs in each group is indicated.}
  \label{fig:pairs}%
\end{figure}

In fact the trajectories can be grouped into 6 classes, as shown in
\figref{fig:pairs}. The numbers of pairs ($N$) in each group are
indicated in the inserts, and these range from 13 to 29. As most of
the drifters in each group were deployed on a single day, these
obviously should not be considered as independent realizations.  The
exception was the $N = 17$ group, which were launched on two different
days. Thus the number of independent realizations here is only 7, 1
for each class and 2 for the $N = 17$ group. However, allowing for
some variation in each group, we estimated the degrees of freedom as 3
times this, or 21. As such, we effectively treat each group as a
triplet of drifters.

With so few degrees of freedom, one cannot distinguish the theoretical
curves at the 95\pct confidence level. Thus both the Lundgren and
Richardson PDFs are statistically similar over the initial 10 d, with
all three separations. Only the Rayleigh distribution can be excluded
during this period.

The relative dispersion is indicated by the red curves in the third
row of panels in \figref{fig:pdfr2r4-glad} (the curves are often
obscured by blue curves indicating the dispersion for a lowpass
filtered data set, discussed hereafter).  The dispersion with the
smallest initial separation is close to exponential initially but
increases more slowly after $t = 2$ d. With $r_0 = 1$ km however the
dispersion follows the Richardson prediction over much of the first 20
d. With this separation and $r_0 = 10$ km, the dispersion asymptotes
to linear growth at late times.

The kurtoses are plotted in red in the bottom panels of
\figref{fig:pdfr2r4-glad}.  With $r_0 = 0.15$ km the kurtosis grows
and then oscillates around the Richardson asymptotic limit of 5.6.
With $r_0 = 1$ km, the kurtosis increases more slowly, lying between
the asymptotic limits for the Richardson and Rayleigh regimes, while
the kurtosis is close to the latter limit for most of the period
with $r_0 = 10$ km.

\begin{figure}
  \centerline{%
  \includegraphics[width=.475\textwidth]{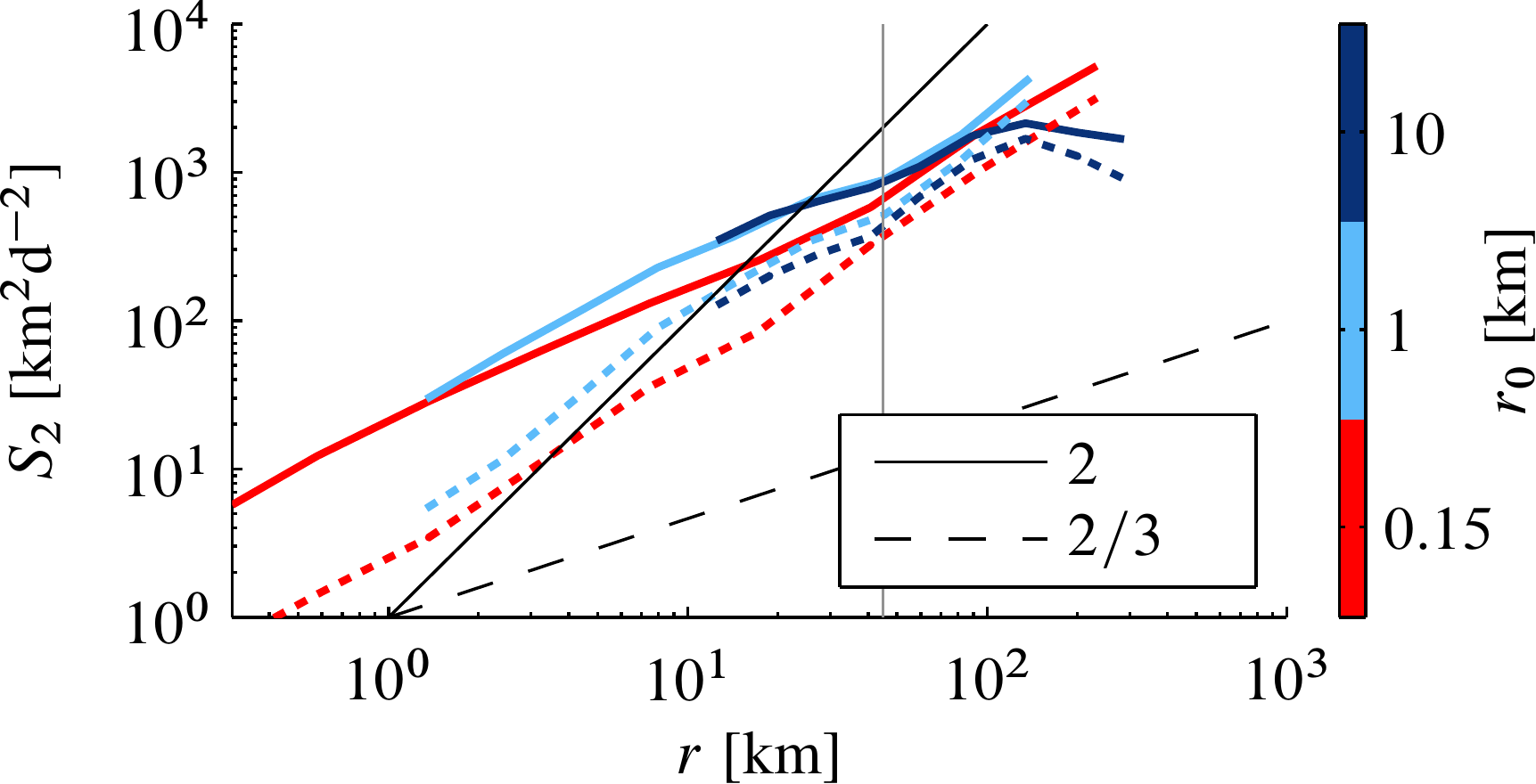}}%
  \caption{As in \figref{fig:s2} but using raw (solid) and lowpassed
  (dashed) GLAD trajectory pairs.}
  \label{fig:s2-glad}%
\end{figure}

The second-order structure functions (\figref{fig:s2-glad}, solid
curves) on the other hand are clearly suggestive of a Richardson
regime.  With the $r_0 = 0.15$ km pairs, the curve exhibits a
power-law-dependence near $r^{2/3}$ from the smallest scales to
roughly 200 km, consistent with the results of \citet{Poje-etal-14}.
However, with $r_0 = 10$ km a plateau is observed above 100 km, in
agreement with the motion being largely uncorrelated at those scales.

Thus the GLAD results are ambiguous. The PDFs are inconclusive, due to
having too few degrees of freedom. The dispersion curves suggest
nonlocal dispersion at small scales while the kurtosis and the
structure functions are more consistent with a Richardson regime. What
causes these differences among the measures, and are the results
actually different than those from the synthetic pairs?

\subsection{Inertial oscillations}

\begin{figure}
  \centerline{%
  \includegraphics[width=.475\textwidth]{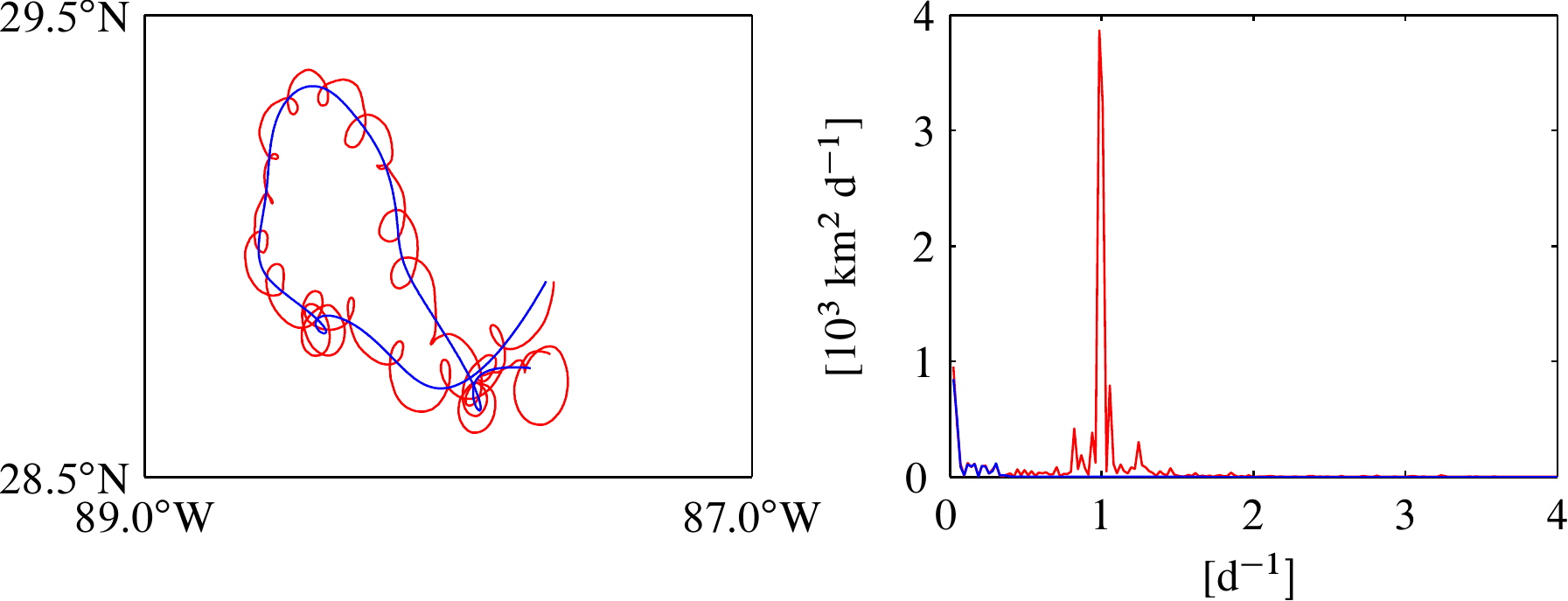}}% 
  \caption{(left) Raw (red) and lowpassed (blue) GLAD drifter
  trajectory.  (right) Corresponding zonal velocity power
  spectra.}
  \label{fig:Puu}%
\end{figure}

One important difference can be seen in the trajectories themselves
(\figref{fig:pairs}). Many of the GLAD drifters experience inertial
oscillations: anticyclonic loops with a period near 1 d (the local
inertial period) \cite[cf., e.g.,][]{Gill-82}.  Frequency spectra
of the individual velocities, as in the example in \figref{fig:Puu},
exhibit a significant peak at the inertial period (right).  Except
for this, the energy resides primarily at the lowest frequencies.
Consistently, the trajectory (left) exhibits anticyclonic loops
superimposed on a larger scale structure.

While inertial oscillations have a narrow frequency spectrum [unless
modified by background rotation \citep[e.g.,][]{Kunze-85}], the
Lagrangian motion possesses a range of spatial scales. These are
determined by the particle velocity, with faster-moving particles
executing larger loops. So the spectral profile in wavenumber space
is broader; as such, the oscillations could potentially influence the
separation statistics.

To test this, we applied a lowpass filter to remove the inertial
oscillations.\footnote{The filter was a sixth-order lowpass Butterworth
filter with 2.5 d$^{-1}$ cut-off frequency.} The resulting spectrum
and the corresponding trajectory are shown in blue in \figref{fig:Puu}.
The filter eliminates the peak near the inertial frequency while
preserving the larger meandering motion.

Lowpass filtering has relatively little effect on the time-based
dispersion measures. The blue curves in \figref{fig:pdfr2r4-glad}
correspond to the filtered GLAD trajectories, and in all cases these
mirror the results for the unfiltered trajectories. This is because
the inertial oscillations on nearby drifters are very similar (see
below) and thus do not greatly alter their separation. Moreover, as
the drifters return approximately to their previous positions every
inertial period, the net effect is small.

The impact on the velocity structure function however is greater
(\figref{fig:s2-glad}). With the lowpass filter (dashed curves),
the energy at small scales is reduced and the structure function
increases faster than $r^{2/3}$.

Inertial oscillations are only weakly captured in the NCOM simulation,
so this effect is missed in the synthetic trajectories, which are
furthermore computed from daily velocities.  To test how they would
have altered the statistics, we \emph{added} inertial oscillations
to the NCOM trajectories. This was done by modifying the $i \neq
j$ positions as
\begin{align}
  x_i(t) &\mapsto x_i(t) + A (\sin\omega t, \cos\omega t - 1)\\
  x_j(t) &\mapsto x_j(t) + \left(A + B(r_{ij})\right) (\sin\omega t, \cos\omega t - 1).
  \label{eq:nio}
\end{align}
Here $2\pi/\omega = 1$ d, roughly equal to the local inertial period.
The amplitude, $A$, was taken to be a random number varying over
the range of observed loop amplitudes. The amplitude $B(r_{ij}(t))$
represents the growing difference between amplitudes on nearby
drifters.

\begin{figure}
  \centerline{%
  \includegraphics[width=.475\textwidth]{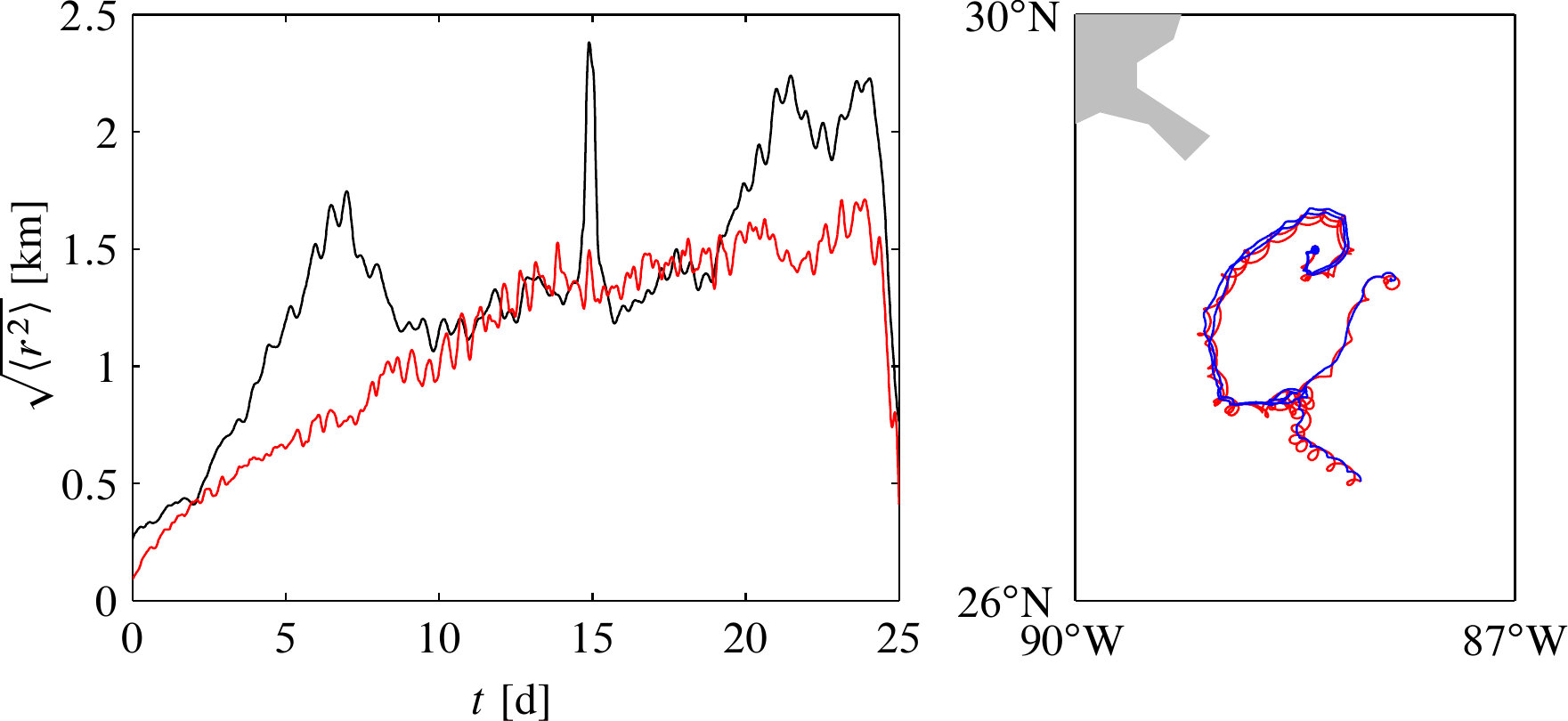}}%
  \caption{(left) The root-mean-squared (rms) separation for
  highpass-filtered GLAD trajectory pairs (black) plotted against
  the rms separation of the lowpass-filtered trajectories.  The red
  curve is an empirical fit used to determine the difference in
  amplitude of the artificial near-inertial oscillations imposed
  on the NCOM trajectories.  The GLAD trajectories belong to the
  $r_0 \approx 1$ km initial separation class and the NCOM trajectories
  start at GLAD positions.  (right) An example of an NCOM pair,
  without (blue) and with (red) the near-inertial oscillations
  superimposed.}
  \label{fig:nio-r2pair}%
\end{figure}

The latter was chosen to mimic the behavior of the GLAD pairs.  The
inertial wave scale is generally much larger than the smallest pair
separations here \citep[e.g.,][]{Webster-68, Pollard-80, Young-Jelloul-97,
Chant-01}.  Consistently, the high-passed pair velocities decorrelate
on scales comparable to those of the full velocities (not shown),
indicating scales of at least 100 km. As the pair velocities are
similar for nearby drifters, the loops are correlated and similarly
large.  But as the drifters separate, the difference in loop amplitude
grows as the velocity difference grows.  To gauge this, we calculated
the rms difference in the highpass filtered pair separation and
plotted it against the rms pair separation, determined from the
lowpass filtered trajectories.  This is shown in black in the left
panel of \figref{fig:nio-r2pair}.  The rms difference is roughly
0.5 km initially and increases to a value near 2 km at separations
greater than 20 km.

Choosing $B(r_{ij}(t)) = B_0\sqrt[4]{r_{ij}(t)/r_0}$ with $B_0 =
3/4$ yields the separation curve shown in red. This mimics the
observed growth fairly well. The effect of the addition on a single
pair is shown in the right panel, with the modified trajectories
exhibiting anticyclonic loops.

\begin{figure}
  \centerline{%
  \includegraphics[width=.475\textwidth]{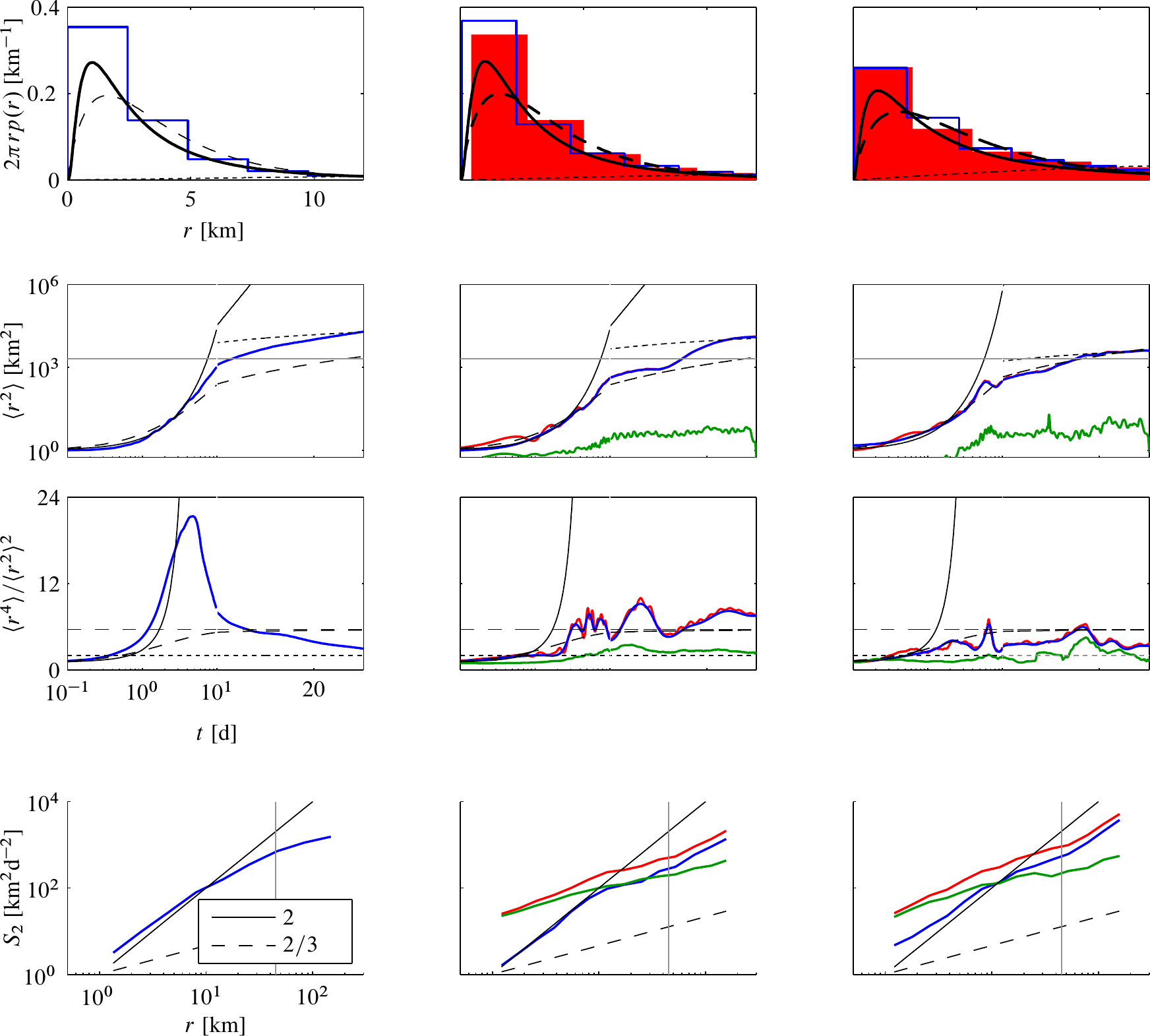}}% 
  \caption{Separation PDFs (top), relative dispersion (second row),
  kurtosis (third row), and velocity structure function (bottom)
  for: NCOM pairs at $r_0 = 1$ km separation, as shown in Figs.\
  \ref{fig:pdf}--\ref{fig:s2}, for comparison (left);  NCOM
  trajectories started from the GLAD positions with $r_0 \approx
  1$ km (middle);  and actual GLAD trajectories with $r_0 \approx
  1$ km  (right).  In blue are results based on NCOM trajectories
  or lowpassed GLAD trajectories.  In red are results based on NCOM
  trajectories with near-inertial oscillations superimposed or raw
  GLAD trajectories.  In green are results based on highpassed
  trajectories.}
  \label{fig:nio-pdfr2r4s2}%
\end{figure}

The effect on the synthetic particle statistics is seen in the
middle panels of \figref{fig:nio-pdfr2r4s2}. These are for a set
of NCOM trajectories initiated at the GLAD launch positions, with
$r_0 = 1$ km.  The blue curves are for the unmodified trajectories
and the red have the inertial oscillations superimposed.  The PDFs,
relative dispersion, and kurtosis are almost unaffected by the
addition of the oscillations. But the structure function is
significantly altered, with that for the modified trajectories
exhibiting more energy at subdeformation scales. While $S_2$ increases
as $r^2$ for the original trajectories, the dependence for the
modified set is nearer $r^{2/3}$.

Shown in green are the statistics for the inertial oscillations
alone, obtained from the trajectories following highpass filtering
(with a cut-off of 2.5 d). These show that the contributions to the
dispersion and kurtosis from the inertial oscillations are much
less than from the low frequency components. But below separations
of roughly 20 km, the inertial component dominates the second order
structure function.

We find that adding the oscillations affects other distance-based
measures as well, such as the relative diffusivity and the FSLE
(not shown).  But as they do not significantly alter the time-based
metrics, they cannot explain the other differences between the GLAD
and NCOM statistics.

\subsection{Sampling}

The second factor influencing the GLAD results is the sampling
strategy alluded to earlier. Most of the GLAD pairs were deployed
very near one another and behaved similarly. This reduced the
effective degrees of freedom, and the mesoscale dispersion was
not captured well.

The effect can be assessed by comparing in \figref{fig:nio-pdfr2r4s2}
the statistics for the $r_0 = 1$ km NCOM pairs deployed at the GLAD
locations (middle) with those from the full NCOM deployment (left)
and the GLAD set (right).  The statistics for the reduced NCOM set
are strikingly similar to those from GLAD, both the time- and the
distance-based measures.  The kurtoses are somewhat lower with the
GLAD drifters, and the transition scale below which the structure
function from the highpassed velocities crosses that from the
lowpassed velocities is smaller (10 vs 20 km). But the addition or
removal of inertial oscillations otherwise affects the trajectories
in the same ways.  And the results from both sets differ markedly
from those for the full NCOM set.

This implies the differences between the GLAD statistics and those
of the full NCOM set may be due to sampling rather than to dynamics.
With identical sampling, the NCOM model gives an accurate representation
of the dispersion and the structure function. This is remarkable,
given that the model resolution is only 1 km.  The reason however
can be inferred from the frequency spectrum in \figref{fig:Puu}:
except for the peak at the inertial frequency, the spectrum is
dominated by low frequency motions. It is these motions, due to
mesoscale features, which dominate the pair dispersion, and these
are captured by the model. This is the essence of nonlocal
dispersion---it is controlled by structures with larger scales and
lower frequencies.

\section{Summary and concluding remarks}

We have investigated the dispersion experienced by simulated and
observed pairs initiated near the \emph{Deepwater Horizon} (DwH)
site in the northern Gulf of Mexico (GoM).  The simulated separations
were produced using synthetic pairs advected by surface velocities
from a data-assimilative Navy Coastal Ocean Model (NCOM) simulation,
with an effective horizontal resolution of 1 km, during July 2013
and February 2014.  The observed separations come from drifter pairs
from the Grand LAgrangian Deployment (GLAD), conducted in July 2012.
The investigation involved various statistical descriptors, namely,
the probability distribution function (PDF) of pair separations,
its second moment (relative dispersion), its fourth moment (kurtosis),
and the (second-order) velocity structure function.

The measures are consistent for the NCOM pairs, suggesting nonlocal
dispersion at the smallest separations and diffusive dispersion at
separations greater than 100 km, where the pair velocities are
decorrelated. Similar results were obtained in both winter and summer
seasons. The results were more ambiguous with the GLAD pairs.  While
the pair motion is also uncorrelated at scales exceeding 100 km, the
dispersion regime cannot be distinguished from the PDFs, dispersion or
kurtosis. The structure functions on the other hand indicate
Richardson dispersion from the smallest scales to beyond the
decorrelation scale.

Two effects impact the GLAD results. With such high temporal
resolution (10-min sampling), the drifters resolve inertial
oscillations. These energetic motions however only weakly affect pair
dispersion.  The size of the loops depend on the drifter velocity
\cite[e.g.,][]{Gill-82}, and since nearby drifters have nearly the
same velocity, the size of the loops is nearly the same initially.  As
the pair separates however, the velocity difference grows and the
amplitude difference increases. However, as the drifters return
approximately to their starting position after an inertial period, the
net effect on the separation during that time is small. As such, the
effect is greatly overwhelmed by mesoscale stirring.

The oscillations do however alter distance-averaged measures. The
velocity structure function is one such measure, and this is dominated
by inertial oscillations below roughly 10 km.  Filtering the
trajectories to remove the inertial oscillations steepens the
structure functions without affecting the time-based measures.
Likewise, adding inertial oscillations to the trajectories from the
model, which has weak inertial variability, causes the structure
function to shallow, lending the appearance of local dispersion.

The second effect concerns the sampling in GLAD. As the goal was
to resolve submesoscale dispersion in the region, the drifters were
deployed in several tight clusters. As these spanned scales much
less than the correlation length scale (100 km), the pairs behaved
similarly. We found the drifters could be separated into 6 distinct
classes, each displaying a characteristic path. This reduced the
degrees of freedom and led effectively to an undersampling of the
mesoscale stirring.  Using synthetic particles deployed at the same
locations yielded nearly identical, and equally ambiguous, dispersion
statistics.

The conclusion is that the 1-km model likely captures the dispersion
in the GLAD experiment over the sampled scales. This argues in favor
of nonlocal dispersion, because the stirring is dominated by large
scale, low frequency motions which are well-resolved by the model.
It also supports using altimeter-derived geostrophic velocities to
study dispersion here, as the dominant eddies are marginally resolved
by altimetry. \citet{Olascoaga-etal-13} suggested the mesoscale
circulation dominates in shaping the patterns formed by drifters
in the GLAD experiment, and the present results are consistent with
this.

Previously, \citet{LaCasce-Ohlmann-03} observed exponential dispersion
among ``chance pairs'' from the Surface-CUrrent and Lagrangian
drifter Program (SCULP), from separations of 1 km up to $L_\mathrm{D}$
($\approx 50$ km).  The SCULP pairs exhibited an e-folding time on
the order of 1 d, similar to the time scales inferred here, and
exhibited large kurtoses. \cite{LaCasce-10} found moreover that the
SCULP PDFs resemble the Lundgren distribution.  Note the SCULP
drifters had daily positions and so were essentially devoid of
inertial oscillations.  \citet{LaCasce-Ohlmann-03} did not observe
diffusive dispersion at super-deformation scales but something
closer to ballistic growth, with the dispersion increasing as
$t^{2.2}$. Given that the pair motion is uncorrelated above
$L_\mathrm{D}$, such growth most likely reflects shear dispersion,
due to a large scale flow. Indeed, many of the SCULP drifters were
advected by boundary currents.

\citet{Poje-etal-14} presented the first analysis of the GLAD pair
trajectories and concluded the dispersion was consistent with the
Richardson regime, from the smallest sampled scale (0.1 km) to
several hundred kilometers.  However, their conclusions were based
solely on distance-based measures (the second order structure
function and the relative diffusivity) which are affected by inertial
oscillations.  It should be emphasized too that a turbulence framework
cannot be applied to interpret results at separations of hundreds
of kilometers, as the pair velocities are uncorrelated.

\citet{Jullien-etal-99} and \citet{Jullien-03} calculated separation
PDFs from pairs of particles deployed in 2D turbulent flows in the
laboratory. They suggested that the separation PDF could be fit
with an empirical function of the form:
\begin{equation}
  p(r,t) = \frac{a}{2\pi\sigma r} \exp\left(-b
  \sqrt{\frac{r}{\sigma}}\right),
  \label{eq:jull}
\end{equation}
where $\sigma = \langle r^2 \rangle^{1/2}$, and $a$ and $b$ are
constants. They claimed that the same PDF applied for
\emph{both} the energy and enstrophy cascade ranges, with slightly
different values of $a$ and $b$.\footnote{The PDF they proposed was not
properly normalized.  Doing so yields $a = b^{2}/2$.} 
It is straightforward to show that the kurtosis for this empirical
PDF \eqref{eq:jull} is:
\begin{equation}
  \frac{\langle r^{4}\rangle }{\langle r^{2}\rangle ^{2}} =
  \frac{9!}{5!^2} = 25.2.
  \label{eq:jullku}
\end{equation}
There is no indication of such a large asymptotic limit in either the
simulated trajectories or the GLAD data.  So we can most likely rule
out this type of dispersion.

The present results serve as a cautionary note on using relative
dispersion to deduce kinetic energy spectra. The inertial oscillations
contribute to the spectra but do not greatly impact dispersion.  So
finding exponential relative dispersion does not necessarily imply
steep spectra. Conversely, having shallower spectra at small scales,
as in the atmosphere \citep{Nastrom-Gage-85} and ocean
\citep{Callies-Ferrari-13}, does not rule out nonlocal pair dispersion.

It is also worth noting that the submesoscale spectrum here exhibits a
scale separation. The Lagrangian frequency spectra has peaks at low
frequencies and at the inertial frequency, with little energy in
between. So it is possible to partition the flow into mesoscale
turbulence and inertial oscillations. If instead an inverse energy
cascade were occurring from the submesoscales, the intermediate scales
and frequencies would be energetic.

The results also have implications for the design of dispersion
experiments. Care should be taken to ensure sufficient sampling at
scales exceeding those of the energy-containing eddies. Otherwise one
might obtain many similar pair trajectories, with a corresponding loss
of statistical confidence.

\acknowledgments

We thank Angelique Haza, Tamay \"Ozg\"okmen, and Andrew Poje for
discussions on pair-separation statistics and Jonathan Lilly on
inertial oscillations. We are also grateful to three anonymous
reviewers, whose comments led to significant improvements in the
manuscript. The GLAD trajectory pairs were kindly identified by
Angelique, who independently computed kurtosis in the several initial
separation classes considered here.  The GLAD drifter trajectory
dataset is publicly available through the Gulf of Mexico Research
Initiative Information \& Data Cooperative (GRIIDC) at
https://data.gulfresearchinitiative.org  (DOI:10.7266, N7VD6WC8).
The NCOM simulation was produced at the Naval Research Laboratory
and can be obtained from http://\allowbreak dx.doi.org/\allowbreak
10.7266/\allowbreak N7FQ9TJ6, N76Q1V5G, and N72Z13F4.  The work was
supported by the Gulf of Mexico Research Initiative (FJBV) and under
grant 221780 from the Norwegian Research Council (JHL).

\appendix[A] 
\appendixtitle{PDF solutions \label{app:pdf}}

Solutions to the Fokker-Planck equation (\ref{eq:fp}) have been
derived for the turbulent inertial ranges. These assume that all
pairs have the same initial separation, so that $p(r,0)=(2 \pi
r)^{-1} \delta(r-r_0)$. Note that $p$ is normalized, i.e., $\langle
r^0 \rangle = 1$. As noted, the solutions can be obtained via the
Laplace transform.

A scale-independent diffusivity $\kappa_2 = \const$ occurs when the
pair motion is uncorrelated.  One- and two-particle statistics
coincide in such a case, which is consistent with $S_2 = \const$
(indeed, $\langle (v_i - v_j)^2\rangle = 2\langle v_i^2 \rangle$,
which does not depend on scale). The solution to \eqref{eq:fp} is
given by
\begin{equation}
  p(r,t) = \frac{1}{4\pi\kappa_2t} I_0\left(\frac{r_0r}{2\kappa_2t}\right)
  \exp\left(-\frac{r_0^2 + r^2}{4\kappa_2t}\right), 
  \label{eq:pRa}
\end{equation}
where $I_0(\,)$ is a zeroth-order modified Bessel function
\citep{LaCasce-10}. In the long-time ($t \gg \kappa_2^{-1} r$, $r
\gg r_0$) asymptotic limit,
\begin{equation}
  p(r,t) \sim \frac{1}{4\pi\kappa_2t}
  \exp\left(-\frac{r^2}{4\kappa_2t}\right),  
  \label{eq:pRa-long}
\end{equation}
which is proportional to the Rayleigh PDF.  The second (raw) moment (or
relative dispersion) of \eqref{eq:pRa-long},
\begin{equation}
  \langle r^2 \rangle \sim 4\kappa_2t,
  \label{eq:r2Ra-long}
\end{equation}
as expected for a normal diffusive process.  The fourth moment
normalized by the relative dispersion (or kurtosis) of \eqref{eq:pRa-long},
\begin{equation}
  \frac{\langle r^4 \rangle}{\langle r^2 \rangle^2} \sim 2,
  \label{eq:r4Ra-long}
\end{equation}
reflecting the self-similarity of the Rayleigh PDF.

Pair motion is correlated in the turbulent inertial ranges. The
Richardson regime corresponds to the energy inertial range
(both in 3D and 2D), and the correlated motion sustains local
dispersion. With $E \propto k^{-5/3}$, the diffusivity has the form
$\kappa_2= \beta r^{4/3}$ \citep{Richardson-26, Obhukov-41,
Batchelor-50}, with the constant $\beta$ is proportional to the
third root of the energy dissipation rate.  The second-order structure
function, which is the inverse Fourier transform of the kinetic
energy spectrum, is $S_2 \propto r^{2/3}$ \citep{Kolmogorov-41}.

The solution to \eqref{eq:fp} is:
\begin{align}
  p(r,t) = {} & \frac{3}{4\pi\beta t r_0^{2/3}r^{2/3}} I_2\left(
  \frac{9r_0^{1/3}r^{1/3}}{2\beta t}\right)\nonumber\\
  & \times \exp\left( -\frac{9(r_0^{2/3}
  + r^{2/3})}{4\beta t}\right),  
  \label{eq:pRi}
\end{align}
where $I_2(\,)$ is a second-order modified Bessel function
\citep{LaCasce-10}.  In the long-time ($t \gg \beta^{-1} r^{2/3}$,
$r \gg r_0$) asymptotic limit
\begin{equation}
  p(r,t) \sim \left(\frac{3}{2}\right)^5 \frac{1}{\pi(\beta t)^3}
  \exp\left(-\frac{9r^{2/3}}{4\beta t}\right), 
  \label{eq:pRi-long}
\end{equation}
which is the 2D analogue of \citeapos{Richardson-26} solution.  The
relative dispersion associated with \eqref{eq:pRi} is:
\begin{equation}
  \langle r^2 \rangle  = \frac{5!}{2}\left(\frac{4\beta t}{9}\right)^3
  M\left(6,3,\frac{9r_0^{2/3}}{4\beta t}\right)
  \exp\left(-\frac{9r_0^{2/3}}{4\beta t}\right), 
  \label{r2Ri}
\end{equation}
where $M(\,,\,,\,)$ is the Kummer's function \citep{Graff-etal-15};
its long-time asymptotic limit is given by
\begin{equation}
  \langle r^2 \rangle \sim 5.2675 \beta^3t^3.
  \label{r2Ri-long}
\end{equation}
The kurtosis of \eqref{eq:pRi} is
\begin{equation}
    1 \leq \frac{\langle r^4 \rangle}{\langle r^2 \rangle^2} <
	 5.6,
  \label{eq:r4Ri}
\end{equation}
(with the equality holding initially), while that of its long-time
asymptotic limit \eqref{eq:pRi-long} is
\begin{equation}
    \frac{\langle r^4 \rangle}{\langle r^2 \rangle^2} \sim 5.6 \, ,
  \label{eq:r4Ri-long}
\end{equation}
which reflects the self-similarity of the Richardson PDF.  

Finally, in the enstrophy cascade inertial range, with $E \propto
k^{-3}$, the diffusivity is $\kappa_2 = T^{-1}r^2$, where $T$ is
proportional to the inverse cubic root of the enstrophy dissipation
rate \citep{Lin-72}.  The corresponding second-order structure
function is $S_2 \propto r^2$ \citep[e.g.,][]{Bennett-84}.

The solution to \eqref{eq:fp} is given by:
\begin{equation}
  p(r,t) = \frac{1}{4\pi^{3/2}(t/T)^{1/2}r_0^2} \exp\left(-\frac{(\ln
  r/r_0 + 2t/T)^2}{4t/T}\right)
  \label{eq:pLu}
\end{equation}
\citep{Lundgren-81, Bennett-06, LaCasce-10}. The relative dispersion is
\begin{equation}
  \langle r^2 \rangle = r_0^2 \exp \frac{8t}{T},
  \label{eq:r2Lu}
\end{equation}
while the kurtosis is
\begin{equation}
  \frac{\langle r^4 \rangle}{\langle r^2 \rangle^2} = \exp
  \frac{8t}{T}.
  \label{eq:r4Lu}
\end{equation}
The Lundgren PDF \eqref{eq:pLu} is lognormal and thus not
self-similar: it gets more peaked in time, possessing increasingly
long tails (at large scales).  Note that the same PDF and exponential
growth occurs with a kinetic energy spectral slope with $\alpha > 3$
\citep{Bennett-84, Babiano-etal-90}.

\appendix[B] 
\appendixtitle{The NCOM simulation\label{app:ncom}}

Configured for the GoM, the NCOM simulation employs assimilation
and nowcast analyses from NCODA (Navy Coupled Ocean Data Assimilation)
\citep{Cummings-05}.  Forecasts are generated by systems linking
NCODA with regional implementations \citep{Rowley-Mask-14} of NCOM
\citep{Barron-etal-06}.  The model has 1-km horizontal resolution
and was initiated on 15 May 2012 from the then operational global
ocean model Global Ocean Forecast System (GOFS) 2.6 \citep{Barron-etal-07}.
Daily boundary conditions are received from the current operational
GOFS using the HYbrid Coordinate Ocean Model (HYCOM)
\citep{Metzger-etal-09}.  The vertical grid is comprised of 49 total
levels; 34 terrain-following $\sigma$-levels above 550 m and 15
lower $z$-levels. The $\sigma$-coordinate structure has higher
resolution near the surface with the surface layer having 0.5-m
thickness. The simulation uses atmospheric forcing at the sea surface
from COAMPS (Coupled Ocean/Atmosphere Mesoscale Prediction System)
\citep{NRL-97} to generate forecasts of ocean state out to 72 h in
3-h increments. The observational data assimilated in these studies
is provided by NAVOCEANO (Naval Oceanographic Office) and introduced
into NCODA via its ocean data quality control process.  Observations
are three-dimensional variational (3D-Var) assimilated
\citep{Smith-etal-11} in a 24-h update cycle with the first guess
from the prior day NCOM forecast.

\appendix[C] 
\appendixtitle{The GLAD experiment\label{app:glad}}

As part of the GLAD experiment, the Consortium for Advanced Research
on Transport of Hydrocarbon in the Environment (CARTHE) funded by
the BP/Gulf of Mexico Research Initiative deployed more than 300
drifters near the \emph{Deepwater Horizon} site over the period the
period 20--31 July 2012.

Most GLAD drifters followed the CODE (Coastal Ocean Dynamics
Experiment) design \citep{Davis-85}, with a drogue at 1-m depth
that reduces windage and wave motion effects.  With an accuracy of
5 m, the drifter were tracked using the GPS (Global Positioning
System) system, which transmitted positions every 5 to 15 min.
Quarter-hourly drifter trajectory records were obtained from the
raw drifter trajectories treated to remove outliers and fill
occasional gaps, and also lowpass filtered with a 15-min cut-off.

Except for the initial deployment, which consisted of 20 drifters
launched individually on 20 July 2012 over the DeSoto Canyon area,
the deployments were carried out in triplets, with the drifters in
each triplet separated roughly 100 m from each other.  The main
deployments consisted of 2 clusters of 30 triplets arranged in
S-shaped configurations.  One cluster was released on 20 July 2012
centered at (88.1$^{\circ}$, 28.8$^{\circ}$N) and the other cluster
on 20 July 2012 at (87.6$^{\circ}$, 29.2$^{\circ}$N).  Each S-track
spanned an area of approximately 8-km $\times$ 10-km and consisted
of 10 nodes spaced 2- to 4-km apart.  Each node was made up of 3
equilateral triangles with 500-m side.  Another cluster of 10
triplets arranged in a triangular configuration spanning an area
similar to that spanned by S-shaped configurations was launched on
29 July 2012 near (87.5$^{\circ}$, 29.0$^{\circ}$N).  Two additional
clusters with 20 triplets in total were released over 30--31 July
2012 near (89.2$^{\circ}$, 27.8$^{\circ}$N) inside a cyclonic eddy
feature of about 50 km in diameter.

\bibliographystyle{ametsoc2014}
%\bibliography{fot}

\begin{thebibliography}{49}
\providecommand{\natexlab}[1]{#1}
\providecommand{\url}[1]{\texttt{#1}}
\renewcommand{\UrlFont}{\rmfamily}
\providecommand{\urlprefix}{URL }
\expandafter\ifx\csname urlstyle\endcsname\relax
  \providecommand{\doi}[1]{doi:\discretionary{}{}{}#1}\else
  \providecommand{\doi}{doi:\discretionary{}{}{}\begingroup
  \urlstyle{rm}\Url}\fi
\providecommand{\eprint}[2][]{\url{#2}}

\bibitem[{Artale et~al.(1997)Artale, Boffetta, Celani, Cencini,, and
  Vulpiani}]{Artale-etal-97}
Artale, V., G.~Boffetta, A.~Celani, M.~Cencini, and A.~Vulpiani, 1997:
  {Dispersion of passive tracers in closed basins: Beyond the diffusion
  coefficient}. \textit{Phys. Fluids}, \textbf{9~(3162)}, 3162--3171.

\bibitem[{Aurell et~al.(1997)Aurell, Boffetta, Crisianti, Paladin,, and
  Vulpiani}]{Aurell-etal-97}
Aurell, E., G.~Boffetta, A.~Crisianti, G.~Paladin, and A.~Vulpiani, 1997:
  {Predictability in the large: An extension of the concept of Lyapunov
  exponent}. \textit{J. Phys. A: Math. Gen.}, \textbf{30}, 1--26.

\bibitem[{Babiano et~al.(1990)Babiano, Basdevant, {LeRoy},, and
  Sadourny}]{Babiano-etal-90}
Babiano, A., C.~Basdevant, P.~{LeRoy}, and R.~Sadourny, 1990: Relative
  dispersion in two-dimensional turbulence. \textit{J. Fluid Mech.},
  \textbf{214}, 535--557.

\bibitem[{Barron et~al.(2006)Barron, Kara, Martin, Rhodes,, and
  Smedstad}]{Barron-etal-06}
Barron, C.~N., A.~B. Kara, P.~J. Martin, R.~C. Rhodes, and L.~F. Smedstad,
  2006: {Formulation, implementation and examination of vertical coordinate
  choices in the global Navy Coastal Ocean Model (NCOM)}. \textit{Ocean
  Modell.}, \textbf{11}, 347--375.

\bibitem[{Barron et~al.(2007)Barron, Smedstad, Dastugue,, and
  Smedstad}]{Barron-etal-07}
Barron, C.~N., L.~F. Smedstad, J.~M. Dastugue, and O.~M. Smedstad, 2007:
  Evaluation of ocean models using observed and simulated drifter trajectories:
  Impact of sea surface height on synthetic profiles for data assimilation.
  \textit{J. Geophys. Res.}, \textbf{112}, C07\,019, \doi{0.1029/2006JC002982}.

\bibitem[{Batchelor(1950)}]{Batchelor-50}
Batchelor, G.~K., 1950: The application of the similarity theory of turbulence
  to atmospheric diffusion. \textit{Quart. J. Roy. Meteor. Soc.}, \textbf{76},
  133--146.

\bibitem[{Bennett(1984)}]{Bennett-84}
Bennett, A.~F., 1984: {Relative dispersion: Local and nonlocal dynamics}.
  \textit{J. Atmos. Sci.}, \textbf{41}, 1881--1886.

\bibitem[{Bennett(2006)}]{Bennett-06}
Bennett, A.~F., 2006: \textit{Lagrangian fluid dynamics}. Cambridge University,
  Cambridge.

\bibitem[{Callies and Ferrari(2013)Callies, and Ferrari}]{Callies-Ferrari-13}
Callies, J., and R.~Ferrari, 2013: Interpreting energy and tracer spectra of
  upper-ocean turbulence in the submesoscale range (1--200 km).
  \textit{J.Phys.Oceanogr.}, \textbf{43}, 2456--2474.

\bibitem[{Callies et~al.(2014)Callies, Ferrari,, and
  B\"uhler}]{Callies-etal-14}
Callies, J., R.~Ferrari, and O.~B\"uhler, 2014: Transition from geostrophic
  turbulence to inertia--gravity waves in the atmospheric energy spectrum.
  \textit{Proc. Nat. Acad. Sci. USA}, \textbf{111}, 17\,033--17\,038.

\bibitem[{Chant(2001)}]{Chant-01}
Chant, R.~J., 2001: {Evolution of near-inertial waves during an upwelling event
  on the New Jersey inner shelf}. \textit{J.Phys. Oceanogr.}, \textbf{31},
  746--764.

\bibitem[{Charney(1971)}]{Charney-71}
Charney, J.~G., 1971: Geostrophic turbulence. \textit{J. Atmos. Sci.},
  \textbf{28,}, 1087--1093.

\bibitem[{Chelton et~al.(1998)Chelton, {deSzoeke}, Schlax, {El Naggar},, and
  Siwertz}]{Chelton-etal-98}
Chelton, D.~B., R.~A. {deSzoeke}, M.~G. Schlax, K.~{El Naggar}, and N.~Siwertz,
  1998: Geographical variability of the first baroclinic rossby radius of
  deformation. \textit{J. Phys. Oceanogr.}, \textbf{28}, 433--460.

\bibitem[{Coelho et~al.(2015)}]{Coelho-etal-15}
Coelho, E.~F., and Coauthors, 2015: {Ocean current estimation using a
  Multi-Model Ensemble Kalman Filter during the Grand Lagrangian Deployment
  experiment (GLAD)}. \textit{Ocean Modell.}, \textbf{87}, 86--106,
  \doi{10.1016/j.ocemod.2014.11.001}.

\bibitem[{Cummings(2005)}]{Cummings-05}
Cummings, J.~A., 2005: Operational multivariate ocean data assimilation.
  \textit{Q. J. Royal Meteorol. Soc.}, \textbf{131}, 3583--3604.

\bibitem[{Davis(1985)}]{Davis-85}
Davis, R., 1985: Drifter observations of coastal surface currents during
  {CODE}: {T}he method and descriptive view. \textit{J. Geophys. Res.},
  \textbf{90}, 4741--4755.

\bibitem[{Gill(1982)}]{Gill-82}
Gill, A.~E., 1982: \textit{Atmosphere-{O}cean {D}ynamics}. Academic.

\bibitem[{Graff et~al.(2015)Graff, Guttu,, and LaCasce}]{Graff-etal-15}
Graff, L.~S., S.~Guttu, and J.~H. LaCasce, 2015: Relative dispersion in the
  atmosphere from reanalysis winds. \textit{J. Atmos. Sci.}, \textbf{72},
  2769--2785, \doi{10.1175/JAS-D-14-0225.1}.

\bibitem[{Jacobs et~al.(2014)}]{Jacobs-etal-14}
Jacobs, G.~A., and Coauthors, 2014: {Data assimilation considerations for
  improved ocean predictability during the Gulf of Mexico Grand Lagrangian
  Deployment (GLAD)}. \textit{Ocean Modell.}, \textbf{83}, 98--117,
  \doi{10.1016/j.ocemod.2014.09.003}.

\bibitem[{Jullien(2003)}]{Jullien-03}
Jullien, M.-C., 2003: {Dispersion of passive tracers in the direct enstrophy
  cascade: Experimental observations}. \textit{Phys. Fluids}, \textbf{5}, 2228.

\bibitem[{Jullien et~al.(1999)Jullien, Paret,, and Tabeling}]{Jullien-etal-99}
Jullien, M.-C., J.~Paret, and P.~Tabeling, 1999: Richardson pair dispersion in
  two-dimensional turbulence. \textit{Phys. Rev. Lett}, \textbf{82},
  2872--2875.

\bibitem[{Klein(2009)}]{Klein-09}
Klein, P., 2009: The oceanic vertical pump induced by mesoscale and
  submesoscale turbulence. \textit{Ann. Rev. Marine Sci.},
  \textbf{1}, 351--375.

\bibitem[{Kolmogorov(1941)}]{Kolmogorov-41}
Kolmogorov, A.~N., 1941: The local structure of turbulence in incompressible
  viscous fluid for very large reynolds numbers. \textit{Dokl. Akad. Nauk
  SSSR}, \textbf{30}, 9--13.

\bibitem[{Koszalka et~al.(2009)Koszalka, LaCasce,, and
  Orvik}]{Koszalka-etal-09}
Koszalka, I., J.~H. LaCasce, and K.~A. Orvik, 2009: {Relative dispersion in the
  Nordic Seas}. \textit{J. Mar. Res.}, \textbf{67}, 411--433.

\bibitem[{Kraichnan(1966)}]{Kraichnan-66}
Kraichnan, R.~H., 1966: Dispersion of particle pairs in homogeneous turbulence.
  \textit{Phys. Fluids}, \textbf{9}, 1937--1943.

\bibitem[{Kraichnan(1967)}]{Kraichnan-67}
Kraichnan, R.~H., 1967: Inertial ranges in two-dimensional turbulence.
  \textit{Phys. Fluids}, \textbf{10}, 1417--1423.

\bibitem[{Kunze(1985)}]{Kunze-85}
Kunze, E., 1985: Near inertial wave propagation in geostrophic shear.
  \textit{J. Phys. Oceanogr.}, \textbf{15}, 544--565.

\bibitem[{LaCasce(2008)}]{LaCasce-08}
LaCasce, J.~H., 2008: {Statistics from Lagrangian observations}. \textit{Progr.
  Oceanogr.}, \textbf{77}, 1--29.

\bibitem[{LaCasce(2010)}]{LaCasce-10}
LaCasce, J.~H., 2010: Relative displacement probability distribution functions
  from balloons and drifters. \textit{J. Mar. Res.}, \textbf{68}, 433--457.

\bibitem[{{LaCasce} and Ohlmann(2003){LaCasce}, and
  Ohlmann}]{LaCasce-Ohlmann-03}
{LaCasce}, J.~H., and C.~Ohlmann, 2003: {Relative dispersion at the surface of
  the Gulf of Mexico}. \textit{J. Mar. Res.}, \textbf{61}, 285--312.

\bibitem[{Lin(1972)}]{Lin-72}
Lin, J.-T., 1972: Relative dispersion in the enstrophy-cascading inertial range
  of homogeneous two-dimensional turbulence. \textit{J. Atmos. Sci.},
  \textbf{29}, 394--395.

\bibitem[{Lundgren(1981)}]{Lundgren-81}
Lundgren, T.~S., 1981: Turbulent pair dispersion and scalar diffusion.
  \textit{J. Fluid Mech.}, \textbf{111}, 27--57.

\bibitem[{McWilliams(2008)}]{McWilliams-08a}
McWilliams, J.~C., 2008: Fluid dynamics at the margin of rotational control.
  \textit{Environ. Fluid Mech.}, \textbf{8}, 441--449.

\bibitem[{Metzger et~al.(2009)Metzger, Smedstad,, and
  Carroll}]{Metzger-etal-09}
Metzger, E.~J., O.~M. Smedstad, and S.~N. Carroll, 2009: {User's Manual for
  Global Ocean Forecast System (GOFS) Version 3.0 (V3.0)}. NRL Memorandum
  Report, NRL/MR/7320-09-9175.

\bibitem[{Morel and Larcheveque(1974)Morel, and
  Larcheveque}]{Morel-Larcheveque-74}
Morel, P., and M.~Larcheveque, 1974: Relative dispersion of constant-level
  balloons in the 200 mb general circulation. \textit{J. Atmos. Sci.},
  \textbf{31}, 2189--2196.

\bibitem[{Nastrom and Gage(1985)Nastrom, and Gage}]{Nastrom-Gage-85}
Nastrom, G.~D., and K.~S. Gage, 1985: A climatology of atmospheric wavenumber
  spectra of wind and temperature observed by commercial aircraft. \textit{J.
  Atmos. Sci.}, \textbf{42}, 950--960.

\bibitem[{{NRL}(1997)}]{NRL-97}
{NRL}, 1997: {The Naval Research Laboratory's Coupled Ocean/Atmosphere
  Mesoscale Prediction System (COAMPS)}. Mon. Weather Rev. 125, 1414-1430.

\bibitem[{Obhukov(1941)}]{Obhukov-41}
Obhukov, A.~M., 1941: Energy distribution in the spectrum of turbulent flow.
  \textit{Izv. Akad. Nauk. SSR Ser. Geogr. Geo.}, \textbf{5}, 453--466.

\bibitem[{Ohlmann and Niiler(2005)Ohlmann, and Niiler}]{Ohlmann-Niiler-05}
Ohlmann, J.~C., and P.~P. Niiler, 2005: {A two-dimensional response to a
  tropical storm on the Gulf of Mexico shelf}. \textit{Progr. Oceanogr.},
  \textbf{29}, 87--99.

\bibitem[{Olascoaga et~al.(2013)}]{Olascoaga-etal-13}
Olascoaga, M.~J., and Coauthors, 2013: {Drifter motion in the Gulf of Mexico
  constrained by altimetric Lagrangian Coherent Structures}. \textit{Geophys.
  Res. Lett.}, \textbf{40}, 6171--6175, \doi{10.1002/2013GL058624}.

\bibitem[{Poje et~al.(2014)}]{Poje-etal-14}
Poje, A.~C., and Coauthors, 2014: {The nature of surface dispersion near the
  Deepwater Horizon oil spill}. \textit{Proc. Nat. Acad. Sci. USA},
  \textbf{111}, 12\,693--12\,698.

\bibitem[{Pollard(1980)}]{Pollard-80}
Pollard, R.~T., 1980: Properties of near-surface inertial oscillations.
  \textit{J. Phys. Oceanogr.}, \textbf{10}, 385--398.

\bibitem[{{Press} et~al.(2007){Press}, {Teukolsky},, and
  {Vetterling}}]{Press-etal-07}
{Press}, W.~H., S.~A. {Teukolsky}, and W.~T. {Vetterling}, 2007:
  \textit{Numerical Recipes: The Art of Scientific Computing}. 3rd ed.,
  Cambridge University, Cambridge.

\bibitem[{Richardson(1926)}]{Richardson-26}
Richardson, L.~F., 1926: Atmospheric diffusion on a distance-neighbour graph.
  \textit{Proc. R. Soc. Lond. A}, \textbf{110}, 709--737.

\bibitem[{Rowley and Mask(2014)Rowley, and Mask}]{Rowley-Mask-14}
Rowley, C., and A.~Mask, 2014: Regional and coastal prediction with the
  relocatable ocean nowcast/forecast system. \textit{Oceanography},
  \textbf{27}, 3, \doi{10.5670/oceanog.2014.67}.

\bibitem[{Smith et~al.(2011)}]{Smith-etal-11}
Smith, S.~R., and Coauthors, 2011: {Validation Test Report for the Navy Coupled
  Ocean Data Assimilation 3D Variational Analysis (NCODA-VAR) System, Version
  3.43}. NRL Memorandum Report NRL/MR/7320-11-9363.

\bibitem[{Thomas et~al.(2008)Thomas, Tandon,, and Mahadevan}]{Thomas-etal-08}
Thomas, L., A.~Tandon, and A.~Mahadevan, 2008: Submesoscale ocean processes and
  dynamics. \textit{Eddy Resolving Ocean Modeling}, H.~Hecht, and H.~Hasumi,
  Eds., AGU, Washington, D. C., 17--38.

\bibitem[{Webster(1968)}]{Webster-68}
Webster, F., 1968: Observations of inertial-period motions in the deep sea.
  \textit{Rev. Geophys.}, \textbf{6}, 473--490.

\bibitem[{Young and Jelloul(1997)Young, and Jelloul}]{Young-Jelloul-97}
Young, W.~R., and M.~B. Jelloul, 1997: Propagation of near-inertial
  oscillations through a geostrophic flow. \textit{J. Mar Res.}, \textbf{55},
  735--766.

\end{thebibliography}

\end{document}